\documentclass[12pt]{iopart}

\usepackage{graphicx}
\usepackage[numbers,sort&compress]{natbib}
\usepackage{amsmath}
\usepackage{amsfonts}
\usepackage{amssymb}
\usepackage{graphicx}
\usepackage{color}
\usepackage{txfonts}
\usepackage[titletoc]{appendix}
\usepackage[colorlinks={true}]{hyperref}
\hypersetup{citecolor={blue}, filecolor={blue}, linkcolor={blue}, urlcolor={blue}}

\begin{document}

\title{Fractional Advection Diffusion Asymmetry  Equation, derivation, solution and application}

\author{Wanli Wang$^{1,2}$ and Eli Barkai $^2$}
\address{$^1$
Department of Applied Mathematics, Zhejiang University of Technology, Hangzhou 310023, China\\
$^2$
Department of Physics, Advanced Materials and Nanotechnology Institute, Bar Ilan University, Ramat-Gan 52900, Israel}

\ead{\mailto{wanliiwang@163.com;~~~~Eli.Barkai@biu.ac.il}}
\vspace{10pt}
\begin{indented}
\item[]
\end{indented}

\begin{abstract}
The non-Markovian continuous-time random walk model, featuring fat-tailed waiting times and narrow distributed displacements with a non-zero mean, is a well studied model for anomalous diffusion.
Using an analytical approach,  we recently demonstrated how a fractional space advection diffusion asymmetry equation, usually associated with Markovian L{\'e}vy flights,  describes the spreading of a packet of particles. Since we use Gaussian statistics  for jump lengths though fat-tailed distribution of waiting times, the appearance of fractional space derivatives in the kinetic equation demands explanations provided in this manuscript. As applications we analyse the spreading of tracers in two dimensions, breakthrough curves investigated in the field of contamination spreading  in hydrology and first passage time statistics. We present a subordination scheme valid for the case when the mean waiting time is finite and the variance diverges, which is  related to L\'evy statistics for the number of renewals in the process.
\end{abstract}

\section{Introduction}\label{18ctrwsect1}

Continuous-time random walk \cite{Shlesinger1974Asymptotic,Montroll1965Random,Bouchaud1990Anomalous,Metzler2000random,Pinaki2007Universal,Weigel2011Ergodic,Meroz2015toolbox,Felix2013Anomalous,Gradenigo2016Field,Kutner2017continuous,Comolli2018Impact} (CTRW) is a  stochastic jump process for a random walk that jumps instantaneously from one site to another, following a sojourn  period on a site. See also recent works in \cite{Barkai2020Packets,Michelitsch2020biased,Vot2020Continuous,Wang2020Large,Adrian2021Large,Jian2022Strong,Michelitsch2022Asymmetric,Burov2022Exponential,Vitali2022anomalous,Marco2023Role}. For example, it was used to describe dispersive transport in time of flight experiment of charge carriers in disordered systems \cite{Scher1975Anomalous}.
The non-Markovian characteristic of CTRW becomes apparent, especially when the disorder is introduced into the system, as observed in the sense of ensemble average \cite{Kutner2017continuous}. Fractional kinetic equations \cite{Sergio2022Fractional,Metzler2000random,Deng2008Finite,Tankou2022unified,Suleiman2022Anomalous,Yan2022New,Evangelista2023Fractional,Tang2023Variable} are a popular tool used to describe various anomalous phenomena \cite{Vladimir2013Fractional,Eraso2021random}. When the bias of the system is a constant, the fractional time diffusion equation follows \cite{Metzler1999Anomalous,Barkai2001Fractional}
\begin{equation}\label{eeeesssase}
\frac{\partial}{\partial t}\mathcal{P}(x,t)=~\!_0\mathcal{D}_t^{1-\alpha}\left[\frac{\partial^2}{\partial x^2}-\frac{\partial}{\partial x}\right]\mathcal{P}(x,t),
\end{equation}
describing anomalous dynamics with $0<\alpha<1$.
Here we make the assumption that both the generalized mobility and the generalized diffusion coefficient are set to one.
The time Riemann-Liouville operator $~\!_0\mathcal{D}_t^{1-\alpha}$ introduces a convolution integral involving a power-law kernel that decays slowly over time, which is related to memory effects of this nonequilibrium system. In addition, $~\!_0\mathcal{D}_t^{1-\alpha}$ is directly governed by heavy-tailed power law waiting times distributions
\begin{equation}\label{powerlawAs}
\phi(\tau)\propto \tau^{-\alpha-1},~~~~\tau\to\infty,
\end{equation}
in the context of the underlying picture, the CTRW model.


Due to heavy-tailed waiting times, characterized by an infinite mean, Eq.~\eqref{eeeesssase} illustrates sub-diffusion behaviors. However, our focus is on the super-diffusion scenario. Our recent findings demonstrated that the positional distribution,  $\mathcal{P}(x,t)$, of the mentioned  CTRW model follows the fractional advection diffusion asymmetry equation (FADAE)
\begin{equation}\label{aaeqfk104}
 \frac{\partial}{\partial t}\mathcal{P}=D\frac{\partial^2}{\partial x^2}\mathcal{P}-V\frac{\partial}{\partial x}\mathcal{P}+S\!~\frac{\partial^\alpha}{\partial(-x)^\alpha}\mathcal{P}
\end{equation}
for $1<\alpha<2$.
\begin{figure} [t!]
	\centering
		\includegraphics[width=0.45\textwidth]{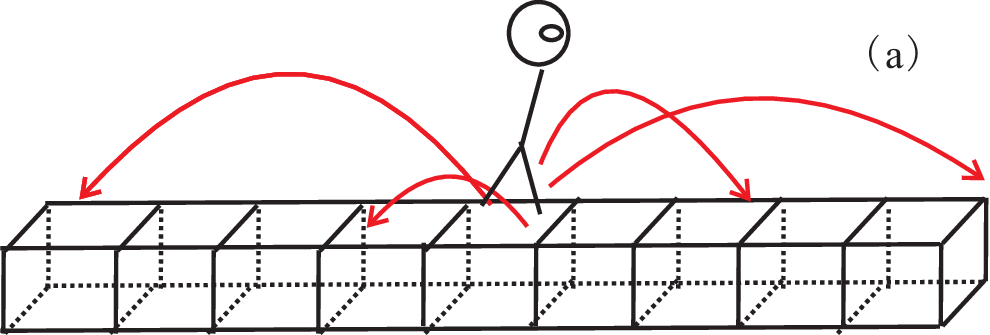}
	\\
		\includegraphics[width=0.45\textwidth]{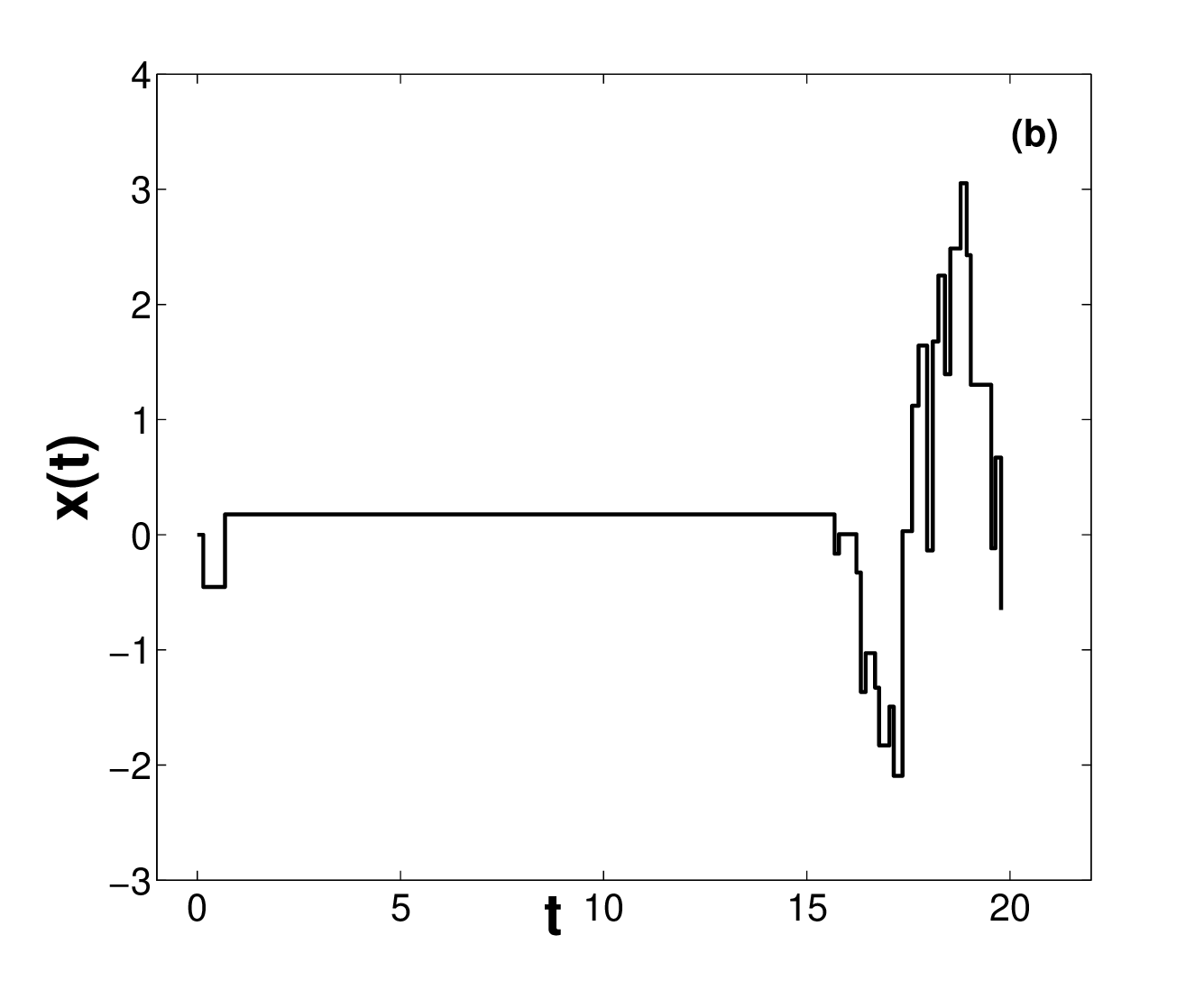}
	\caption{A random walker moves on a one-dimensional lattice with fat-tailed displacements that describe statistics of L{\' e}vy flights [see subplot (a)]. While, subplot (b) displays a trajectory of a CTRW walker with thin tailed displacements, e.g. Gaussian, but the waiting time probability density functions (PDFs) are fat-tailed. Under certain conditions, the packets of spreading particles can appear the same, despite the significant differences of the underlying paths.}
\label{bigjump}
\end{figure}
Here $D$, $V$, and $S$ are transport coefficients to be discussed, and the operator $\frac{\partial^\alpha}{\partial(-x)^\alpha}$  is the right Riemann-Liouville derivative \cite{Oldham1974Fractional,Benson2000fractional-order} in regard to space.  Thus the fat tailed distribution of sojourn times Eq.~\eqref{powerlawAs}, which leads to the fractional time Fokker-Planck equation \eqref{eeeesssase}, when $0<\alpha<1$, can lead as shown here to the fractional space differential equation \eqref{aaeqfk104} if $1<\alpha<2$. There is a transition in the basic kinetic description of the transport.
When the mean waiting time diverges, we get the mentioned fractional time Fokker-Planck equation. However, if the mean sojourn time is finite but the variance diverges, we obtain a fractional space description.
Our first goal is to clarify this point.

Fractional space transport equations \cite{Oldham1974Fractional,Samko1993Fractional,Podlubny1999Fractional,Metzler2000random,Meerschaert2012Stochastic,Vladimir2013Fractional,Estrada2019Space,Zheng2020Error} are usually associated with L{\' e}vy flights \cite{Benson2000fractional-order,Hufnagel2004Forecast,Michael2005Optimal,brockmann2006scaling,Benson2001Fractional,Zaburdaev2015Levy,Yong2016Backward,Guo2021heat,sin2023cauchy}. See subplot (a) in Fig.~\ref{bigjump}. The term L{\' e}vy flights, also referred to as L{\' e}vy motion, was coined by Benoit Mandelbrot in honour of the French mathematician Paul L{\' e}vy. The individual jump has a length
that is distributed with the PDF decaying at large positions as a fat-tailed jump length distribution with diverging variance of the size of the jumps.
Importantly, the L\'evy flight process is a Markovian process, hence the appearance of the fractional framework  in the context of a model with fat tailed distributions of waiting times is of interest. In particular, any fitting of data using the fractional space kinetic equation, cannot be used as strong evidence for L{\'e}vy flights, or for a basic Markovian  underlying process.

Subordination methods, based on an integral transformation,  present a way of solving fractional kinetic equations \cite{Bouchaud1990Anomalous,Saichev1997Fractional,Gorenflo2007Continue,Karina2010Overshooting,Dybiec2010Subordinated,Wang2020Fractional,Chechkin2021Relation,Zhou2022Generalized,Wang2022Ergodic,Lucianno2023Non}.
To be more exact, when the mean waiting time diverges \cite{Barkai2001Fractional}, subordination methods map a classical Fokker-Planck equation onto a fractional diffusion one with the fractional time derivative.
For this case, the term "inverse L{\'e}vy transform"   is sometimes used due to the number of renewals that follows the inverse one-sided L{\'e}vy distribution. This well-known framework works when $0<\alpha<1$. We will propose a new subordination method to solve a wide range of problems valid when the mean waiting time is finite, but the variance is diverging, namely the case $1<\alpha<2$.

%
%
%
%

%
%
%
%
%
%
%
%
%

The paper is  organized as follows.  We start by introducing the CTRW model and give the corresponding concepts in Sec. \ref{18sect2}. We derive the corresponding FADAE and give our explanations in Sec.~\ref{18ctrwSect5}. In Sec.~\ref{applications}, the applications and extensions of Eq.~\eqref{aaeqfk104} are considered, ranging from the FADAE in two dimensions, the positional distribution with the time-dependent bias and variance, breakthrough curves, and the first passage time obtained from subordination methods.  Three limiting laws of FADAE, describing the large asymmetric parameter $S$, the large diffusion term $D$, and the general typical fluctuations,  are discussed  in Sec.~\ref{18ctrwsect2}.  To conclude, we summarize our findings in Sec.~\ref{18ctrwSect8}.

\section{CTRW model}\label{18sect2}
Now we define the CTRW model \cite{Metzler2000random,Berkowitz2002Physical,Kutner2017continuous} discussed in this manuscript. Let us consider the process starting at $t=0$  with the initial position $x=0$.
A walker is trapped on the origin for time $\tau_1$, then makes a jump and the displacement is $\chi_1$; the walker is further trapped on $\chi_1$ for time $\tau_2$, and then jumps to a new position; this process is then repeated. Thus, the process is characterized by a set of waiting times $\{\tau_1, \tau_2, \cdots, \tau_N, B_t\}$ and  the displacements $\{\chi_1, \chi_2, \cdots, \chi_N\}$, where $B_t$ is the backward recurrence time \cite{Godreche2001Statistics,Wang2018Renewal} and  the time dependent  $N$ is the number of renewals from the observation time zero to $t$. Specifically, $\sum_{i=1}^N\tau_i+B_t=t$. These random variables, i.e., $\{\tau_1, \tau_2, \cdots, \tau_N \}$ and $\{\chi_1, \chi_2, \cdots, \chi_N\}$   are mutually independent and identically distributed (IID)  random variables  with common  PDFs $\phi(\tau)$ and $f(\chi)$, respectively.
Consider the broad distribution characterized by a fat tail
\begin{equation}\label{ldeq3201111}
\displaystyle \phi(\tau)=\left\{
          \begin{array}{ll}
            0, & \hbox{$\tau<\tau_0$;} \\
            \alpha\frac{\tau_0^\alpha}{\tau^{1+\alpha}}, & \hbox{$\tau\geq\tau_0$,}
          \end{array}
        \right.
\end{equation}
where $\tau_0$ is a  time scale and the power-law index $1<\alpha<2$. This indicates that the mean $\langle\tau\rangle=\int_0^\infty\tau\phi(\tau)d\tau=\alpha\tau_0/(\alpha-1)$ is finite, but not the variance. The Laplace transform will be used in solving our problems. It is defined by $\widehat{g}(s)=\int_0^{+\infty}\exp(-s\tau)g(\tau)d\tau$ for a well behaved function $g(\tau)$. The Laplace  transform of $\phi(\tau)$, $\tau\to s$, is \cite{Metzler2000random,Godreche2001Statistics}
\begin{equation}\label{laplacepowerlaw}
\widehat{\phi}(s)\sim 1-\langle\tau\rangle s+b_\alpha s^\alpha
\end{equation}
with $b_\alpha=(\tau_0)^\alpha|\Gamma(1-\alpha)|$.
For $s\to 0$, we can check that $\widehat{\phi}(0)=0$, which indicates that $\phi(\tau)$  in
Eq.~\eqref{ldeq3201111} is normalized.

For the displacement PDF $f(\chi)$, we assume $\chi$ has a finite mean $a>0$ and the variance $\sigma^2$. In this manuscript, we consider a Gaussian distribution
\begin{equation}\label{18eq104}
  f(\chi)=\frac{1}{\sqrt{2\sigma^2\pi}}\exp\left[-\frac{(\chi-a)^2}{2\sigma^2}\right].
\end{equation}
In Fourier space,
  $\widetilde{f}(k)=\exp(-\sigma^2 k^2/2-ika)$ with $\widetilde{f}(k)=\int_{-\infty}^\infty \exp(-ikx)f(x)dx$. Thus, the small $k$ expansion of $\widetilde{f}(k)$ reads
\begin{equation}\label{18eq105}
\widetilde{f}(k)\sim1-ika-\frac{1}{2}(\sigma^2+a^2)k^2.
\end{equation}
When the sojourn time PDF has a fat  tail and the variance of the displacement is finite, a wide range of anomalous behaviors emerge \cite{Bouchaud1990Anomalous,Metzler2014Anomalous}. As mentioned, here we will focus on the widely applicable
case $1<\alpha<2$ \cite{Margolin2002Spatial,Levy2003Measurement,Schroer2013Anomalous,Liang2015Sample,Alon2017Time,Alessandro2019Single,Wang2019Transport,Zhang2022Correlated}.



\section{Fractional-Space Advection Diffusion Asymmetry Equation}\label{18ctrwSect5}

\subsection{Calculation of the Positional distribution }
%
%
We now consider the CTRW model and note that  the PDF of the position of a walker at time $t$ is
\begin{equation}\label{18eq500}
P(x,t)_{\rm {CTRW}}=\sum_{N=0}^{\infty}Q_t(N)\chi_N(x)\to\int_{0}^\infty Q_t(N)\chi_N(x){\rm d}N,
\end{equation}
where $Q_t(N)$ denotes the probability of the number of events during the time interval $(0,t)$  and $\chi_N(x)$ is the probability that the walker is on $x$ conditioned that it made $N$ steps. Equation~\ref{18eq500} is  known as the subordination of the spatial process $x$ by the temporal process for $N$ \cite{Bouchaud1990Anomalous,Barkai2001Fractional,Fogedby1994Langevin,Chechkin2021Relation}. The technique was used in computer simulation in the context of fractional Fokker-Planck dynamics \cite{Marcin2007Fractional}, random diffusivity \cite{Wang2022Ergodic}, population heterogeneity \cite{Sergei2023Population} and  one-dimensional Brownian search problem \cite{Zhou2022Generalized}.
As mentioned before, time intervals and displacements of walkers under investigation are IID with common  PDFs $\phi(\tau)$ and $f(\chi)$, respectively.
Note the relation between $\chi_{N-1}(x)$ and $\chi_{N}(x)$, i.e., $\chi_{N}(x)=\int_{-\infty}^{\infty}\chi_{N-1}(y)f(x-y){\rm d}y$, and the density of the particle's displacement after the first step is given exactly by $f(x)$, i.e., $\chi_1(x)=f(x)$. Then  using  the convolution of the Fourier transform, it follows that as well known
\begin{equation}\label{GaussianN}
\chi_N(x)=\frac{1}{\sqrt{2\pi\sigma^2 N}}\exp\left(-\frac{(x-aN)^2}{2\sigma^2 N}\right).
\end{equation}
Based on the renewal process \cite{Godreche2001Statistics}, in Laplace space $\widehat{Q}_s(N)$ follows $\widehat{Q}_s(N)=\widehat{\phi}(s)^N(1-\widehat{\phi}(s))/s$. Considering the random variable $\epsilon=N-t/\langle\tau\rangle$ and taking the inverse transform, then typical fluctuations follow \cite{Kotulski1995Asymptotic,Godreche2001Statistics,Wang2018Renewal}
\begin{equation}\label{18eq501a}
Q_t(N)\sim \frac{1}{(t/\overline{t})^{1/\alpha}}\mathcal{L}_{\alpha}\left(\frac{N-t/\langle\tau\rangle}{(t/\overline{t})^{1/\alpha}}           \right)
\end{equation}
with \begin{equation}\label{sssdhd102}
\overline{t}=\langle\tau\rangle^{1+\alpha}/((\tau_0)^\alpha|\Gamma(1-\alpha)|).
\end{equation}
Eq. \eqref{18eq501a} is valid in the limit of $N-t/\langle\tau\rangle\propto (t/\overline{t})^{1/\alpha}$.  Rare fluctuations  describing  the scaling of $N-t/\langle\tau\rangle\propto t/\langle\tau\rangle$ were investigated in Ref.~\cite{Wang2018Renewal}. Here $\mathcal{L}_{\alpha}(x)$ is the L\'{e}vy  distribution \cite{Oldham1974Fractional,Benson2000fractional-order}, defined by
\begin{equation}\label{deflevystable}
\mathcal{L}_{\alpha}(x)=\frac{1}{2\pi}\int_{-\infty}^{\infty}\exp(ikx)\exp[(-ik)^\alpha]{\rm d}k.
\end{equation}
Utilizing Eqs.~\eqref{18eq500}, \eqref{GaussianN} and \eqref{18eq501a},  we have
\begin{equation}\label{18eq502}
P(x,t)_{\rm {CTRW}}\sim\frac{1}{(t/\overline{t})^{1/\alpha}}\int_{0}^{\infty}\mathcal{L}_{\alpha}\left(\frac{N-t/\langle\tau\rangle}{(t/\overline{t})^{1/\alpha}}\right)\frac{\exp\Big(-\frac{(x-aN)^2}{2\sigma^2 N}\Big)}{\sqrt[]{2\pi\sigma^2 N}}{\rm d}N.
\end{equation}
Note that Eq.~\eqref{18eq502} can be extended to more general situations, e.g. if the jump size distribution is fat-tailed.
Starting from Eq.~\eqref{18eq502} and changing variable $y=(N-t/\langle\tau\rangle)/(t/\overline{t})^{1/\alpha}$,  we get
\begin{equation}\label{18eq503}
P(x,t)_{\rm {CTRW}}\sim\int_{-\frac{t}{\langle\tau\rangle}(\overline{t}/t)^{1/\alpha}}^{\infty}\mathcal{L}_{\alpha}(y)\frac{\exp\Big(-\frac{(x-a\frac{t}{\langle\tau\rangle}-ay(\frac{t}{\overline{t}})^{1/\alpha})^2}{2\sigma^2(t/\langle\tau\rangle+y(t/\overline{t})^{1/\alpha})}\Big)}{\sqrt[]{2\sigma^2\pi(\frac{t}{\langle\tau\rangle}+y(\frac{t}{\overline{t}})^{1/\alpha})}}{\rm d}y,
\end{equation}
describing the scaling when $x-at/\langle\tau\rangle\propto a(t/\bar{t})^{1/\alpha}$.
Note that in the long time limit, the lower limit of the integral in  Eq.~\eqref{18eq503} can be replaced with $-\infty$; see Eq.~\eqref{aaeqfk101} below. The reason is as follows the dimensionless parameter
\begin{equation}\label{18eq505a}
\frac{t}{\langle\tau\rangle}\left({\overline{t}\over t}\right)^{1/\alpha}=\left({t \over \tau_0}\right)^{1-1/\alpha}\left({\alpha \over \Gamma(2-\alpha)}\right)^{1/\alpha}\to \infty
\end{equation}
for $1<\alpha<2$ when $t\to \infty$. The scaling form of the position reads 
\begin{equation}\label{18eq504}
P_{\rm {CTRW}}(\xi,t)\sim\int_{-\frac{t}{\langle\tau\rangle}(\frac{\overline{t}}{t})^{\frac{1}{\alpha}}}^{\infty}\mathcal{L}_{\alpha}(y)\frac{\exp(-\frac{(\xi-y)^2}{2gg(y,t)})}{\sqrt[]{2\pi gg(y,t)}}{\rm d}y
\end{equation}
with
\begin{equation}
\xi=(x-at/\langle\tau\rangle)/(a(t/\overline{t})^{1/\alpha})
\end{equation}
and
\begin{equation}\label{18eq505}
gg(y,t)=
\frac{\sigma^2(t/\langle\tau\rangle+y(t/\overline{t})^{1/\alpha})}{a^2 (t/\overline{t})^{2/\alpha}}.
\end{equation}
In the long time limit, Eq.~\eqref{18eq504} reduces to the well-known result \cite{Kotulski1995Asymptotic,Burioni2013Rare}
\begin{equation}\label{18eq109}
P_{\rm {CTRW}}(\xi)\sim \mathcal{L}_{\alpha}(\xi),
\end{equation}
where we used the fact that $\exp[-z^2/(2\rho^2)]/\sqrt{2\pi \rho^2}$ tends to $\delta(z)$ when $\rho \to 0^{+}$ from Eq.~\eqref{18eq109}.
 More precisely, Eq.~\eqref{18eq109} is valid for the fixed $\sigma$,  $a$, and the long observation time $t$.
At the same time, we  will consider a second dimensionless parameter $\overline{\sigma}^2$,
\begin{equation}\label{sigmaneq}
    gg(y,t) \sim \frac{\sigma^2}{a^2}\left(\frac{t}{\tau_0} \right)^{1-\frac{2}{\alpha}}\left(\frac{\alpha}{\alpha-1}\right)^{\frac{2+\beta}{\beta}}|\Gamma(1-\alpha)|^{-\frac{2}{\alpha}}=\overline{\sigma}^2,
\end{equation}
which quantifies the convergence to the L{\'e}vy distribution.
We will then discuss three limits. If $\overline{\sigma}^2\to 0$, Eq.~\eqref{18eq504} gives Eq.~\eqref{18eq109}.
However, assume a situation when the bias is weak meaning $\sigma^2/a^2$ is very large, this is important in many applications that we consider, and corresponds to the linear response regime when the external driving force determined by $a$ is small. Then the last limit will be to consider a long time but $\bar{\sigma}^2$ finite. In this case, Eq. \eqref{18eq503} is valid. Refer to Appendix \ref{xxxxxB} for more detailed discussions on $\overline{\sigma}^2$.

\begin{figure}[htb]
  \centering
  \includegraphics[width=9cm, height=6cm]{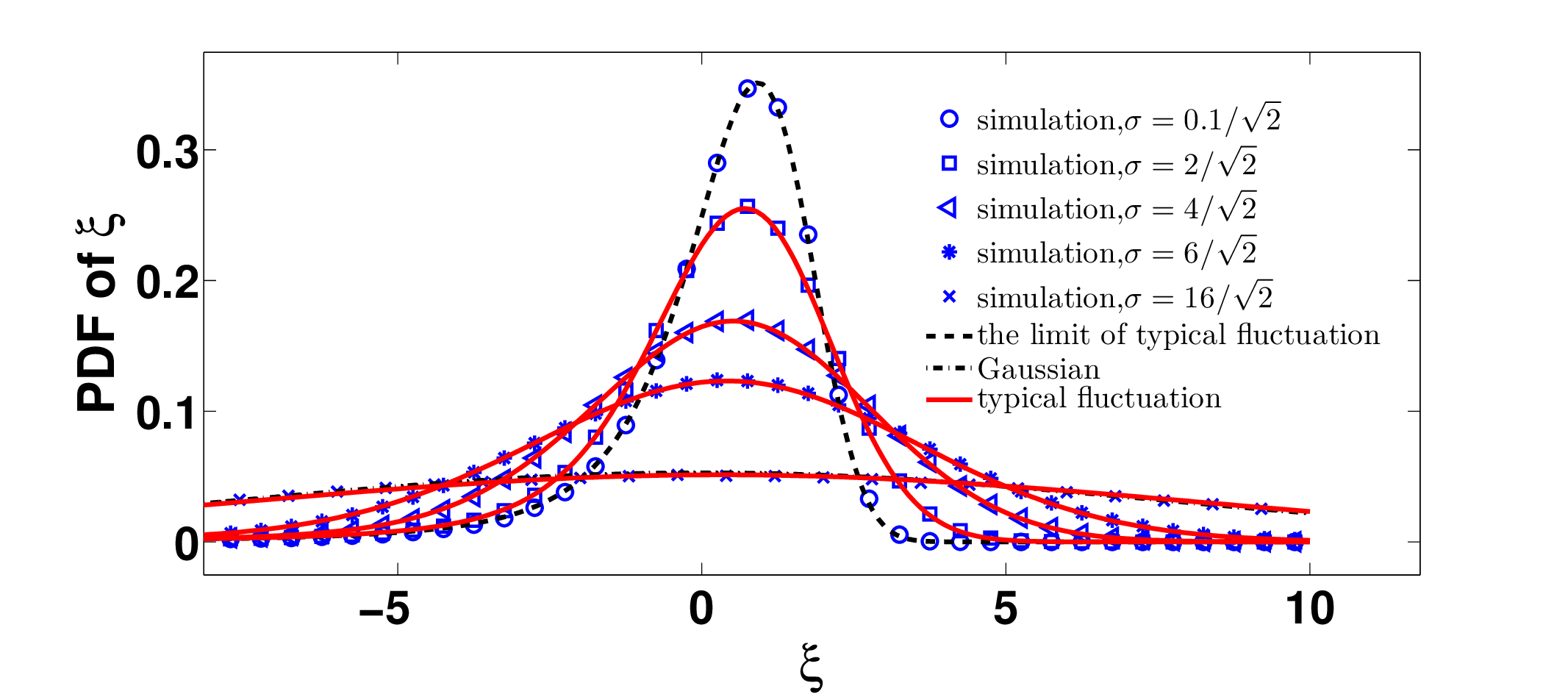}\\
  \caption{PDF of $\xi$ for various $\sigma$ listed in figure. Here the first moment of waiting times, $\langle\tau\rangle=0.3$,  implies that $t/\langle\tau\rangle\approx3333.3\gg 1$. The simulation results, represented by the blue symbols, were obtained by averaging $5\times 10^5$ particles with $\alpha=1.5$ and $t=1000$. The corresponding theoretical prediction, shown by the solid red lines, was derived from Eq.~\eqref{aaeqfk101}, i.e., the exact solution of Eq.~\eqref{aaeqfk104}, and it exhibits a perfect match with no fitting.
  Notice that what the limit theorem Eq.~\eqref{18eq109} predicts here is the top  dashed black line, while our results valid for a variety of  $a$ and $\sigma$ show the deviations from the L{\'e}vy stable distribution $\mathcal{L}_{\alpha}(\xi)$ when $\sigma=2/\sqrt{2},4/\sqrt{2},6/\sqrt{2}$, and $16/\sqrt{2}$. See further discussion in Appendix \ref{xxxxxB}.}
\label{PositionEventTypical}
\end{figure}
\subsection{Derivation of Fractional-Space Advection Diffusion asymmetry Equation}
%

Rewriting  Eq.~\eqref{18eq503} with a lower limit $-\infty$, in the long time limit we get
\begin{equation}\label{aaeqfk101}
\mathcal{P}(x,t)=\int_{-\infty}^{\infty}\mathcal{L}_{\alpha}(y)\frac{\exp\left(-\frac{(x-a\frac{t}{\langle\tau\rangle}-ay(\frac{t}{\overline{t}})^{\frac{1}{\alpha}})^2}{2\sigma^2t/\langle\tau\rangle}\right)}{\sqrt[]{2\sigma^2\pi t/\langle\tau\rangle}}{\rm d}y;
\end{equation}
see Fig. \ref{PositionEventTypical}.
Here we use the expression $\mathcal{P}(x,t)$ to denote the positional distribution $P_{{\rm CTRW}}(x,t)$  in the long time limit.
The statistic of $\mathcal{P}(x,t)$ at a short time $t$  is now discussed. For this case, Eq.~\eqref{aaeqfk101} reduces to
\begin{equation}
\mathcal{P}(x,t)\simeq\frac{\exp(-\frac{x^2}{2\sigma^2 t/\langle\tau\rangle})}{\sqrt{2\pi\sigma^2 t/\langle\tau\rangle}},
\end{equation}
where we ignored the terms $at/\langle\tau\rangle$ and $ay(t/\overline{t})^{1/\alpha}$ in Eq.~\eqref{aaeqfk101}.
It implies that for small $t$ the Gaussian distribution is found. Besides, using $\delta(x)=\lim_{\rho\to 0^{+}}\exp(-x^2/(2\rho^2))/\sqrt{2\pi \rho^2}$, we can see that the initial position of the process is $\mathcal{P}(x,t\to 0)=\delta(x)$.\\
Let us  proceed with the discussion of the FADAE. Taking the Fourier transform of Eq.~\eqref{aaeqfk101}  with respect to $x$  gives
\begin{equation}\label{aaeqfk102}
\displaystyle
\begin{array}{ll}
\mathcal{\widetilde{P}}(k,t)&= \displaystyle \int_{-\infty}^{\infty}\mathcal{L}_{\alpha}(y) \exp\left(-i\left(a\frac{t}{\langle\tau\rangle}+ay\left({t\over\overline{t}}\right)^{1/\alpha}\right)k-k^2\frac{\sigma^2t}{2\langle\tau\rangle}\right)dy
\\
&=\exp\left(-\frac{t\sigma^2}{2\langle\tau\rangle}k^2-iak\frac{t}{\langle\tau\rangle}+(-ik)^\alpha a^\alpha {t\over\overline{t}}\right).
\end{array}
\end{equation}
Let $t$ be zero, Eq.~\eqref{aaeqfk102} results in $\mathcal{\widetilde{P}}(k,t\to 0)=1$, illustrating the fact that $\mathcal{P}(x,t\to 0)=\delta(x)$.
By application of the differential operator $
\partial/\partial t$, we find
FADAE in Fourier space
\begin{equation}\label{aaeqfk103}
\frac{\partial}{\partial t}\mathcal{\widetilde{P}}(k,t)=-\frac{\sigma^2}{2\langle\tau\rangle}k^2\mathcal{\widetilde{P}}(k,t)-ik\frac{a}{\langle\tau\rangle}\mathcal{\widetilde{P}}(k,t)+(-ik)^\alpha\frac{a^\alpha}{\overline{t}}\mathcal{\widetilde{P}}(k,t).
\end{equation}
Taking the inverse Fourier transform, in $(x,t)$ space, we get Eq.~\eqref{aaeqfk104} given in the introduction \cite{Wang2020Fractional}. The transport constants of Eq.~\eqref{aaeqfk104} are as follows
\begin{equation}\label{wqee1021}
 D=\frac{\sigma^2}{2\langle\tau\rangle},~~~~
 V=\frac{a}{\langle\tau\rangle},~~~~
 S=\frac{a^\alpha}{\overline{t}}.
\end{equation}
Here $D$ carrying the dimension $[\sigma^2/\langle\tau\rangle]={\rm cm}^2{\rm s}^{-1}$ is the diffusion constant.
The drift of the process is  $V$. The last constant $S$ on the right-hand side of Eq.~\eqref{aaeqfk104} describes the asymmetric property of the process. 
The operator $\partial^\alpha/\partial(-x)^\alpha$ is the right Riemann-Liouville derivative \cite{Oldham1974Fractional,Benson2000fractional-order} with respect to space  whose Fourier transform is $\mathcal{F}[\frac{\rm{d}^\alpha}{\rm{d}(-x)^\alpha} g(x)]=(-ik)^\alpha\widetilde{g}(k)$. In real space
\begin{equation}\label{Weyl1}
\frac{{\rm d}^\alpha g(x)}{{\rm d} (-x)^\alpha} =\frac{(-1)^n}{\Gamma(n-\alpha)} \frac{{\rm d}^n}{{\rm d} x^n}\int_x^\infty (y-x)^{n-\alpha-1}g(y)\rm{d}y,
\end{equation}
where $n$ is the smallest integer  larger than $\alpha$.  We have used $a>0$ in Eq. \ref{aaeqfk104}, but as discussed below, when $a<0$, the left Riemann-Liouville derivative  comes into being.
We note by passing that mathematically the left Riemann-Liouville  derivative is defined as \cite{Oldham1974Fractional,Benson2000fractional-order}
\begin{equation}\label{Weyl2}
\frac{{\rm d}^\alpha}{{\rm d} x^\alpha}g(x) =\frac{1}{\Gamma(n-\alpha)} \frac{{\rm d}^n}{{\rm d} x^n}\int_{-\infty}^x (x-y)^{n-\alpha-1}g(y)\rm{d}y
\end{equation}
with $\mathcal{F}[\frac{\rm{d}^\alpha}{\rm{d} x^\alpha} g(x)]=(ik)^\alpha\widetilde{g}(k)$.

Here we present a summary of key insights extracted from Eq.~\eqref{aaeqfk104}:
\begin{enumerate}
\item The first two constants, i.e., $D$ and $V$,  are standard relations describing classical advection-diffusion equation. If $a\to 0$, both the asymmetry parameter $S$ and the fractional space derivative become negligible, see Eq. \ref{wqee1021}. The same holds for the drift $V$.
\item Note that the solution $\mathcal{P}(x,t)$ of Eq.~\eqref{aaeqfk104} is normalized for any observation time $t$ since $\mathcal{\widetilde{P}}(k=0,t)=1$ according to Eq.~\eqref{aaeqfk102}.
\item We can check that the exact solution of Eq.~\eqref{aaeqfk104} is Eq.~\eqref{aaeqfk101}. To be more exactly, using Eq.~\eqref{aaeqfk102}, we find that the solution is the convolution of L{\'e}vy distribution and Gaussian distribution, namely
    \begin{equation}\label{shdhshw101}
    \mathcal{P}(x,t)=\frac{\mathcal{L}_{\alpha}\left(\frac{x}{(St)^{\alpha}}\right)}{(St)^{\alpha}}\bigotimes \frac{\exp\left(-{(x-Vt)^2\over4D t}\right)}{\sqrt{4\pi D t}},
    \end{equation}where $\bigotimes$ is the  convolution symbol of the Fourier transform with respect to $x$.

\item When particles are only allowed to move in one direction, for example, $f(\chi)=\delta(\chi-a)$. For this case, $\chi_N(x)$ behaves as $\chi_N(x)=\delta(x-aN)$. Utilizing Eqs.~\eqref{18eq500} and \eqref{18eq501a}, we have $$P_{\rm {CTRW}}(x,t)\sim \frac{1}{a(t/\bar{t})^{1/\alpha}}\mathcal{L}_{\alpha}\left(\frac{\frac{x}{a}-\frac{t}{\langle\tau\rangle}}{(t/\bar{t})^{1/\alpha}}\right).$$
This is the same as Eq.~\eqref{18eq109} after changing variables. Here $\chi_N(x)$ can be treated as a delta function when the variance of displacement is finite. The fractional equation related to this case is given by Eq.~\eqref{aaeqfk104}, where the diffusion term is absent, indicating a strong bias scenario.

  \item Under the influence of the bias,  the fractional derivative operator $\frac{{\rm d}^\alpha}{{\rm d} (-x)^\alpha}$ is with respect to space, though the distribution of the step length is Gaussian. In that sense, the fat-tailed displacement is not a basic requirement for the fractional space derivative operator.

\item The order of taking limits of $t$ approaching infinity and $a$ going to zero cannot be interchanged. As the bias of the system approaches zero, the solution of Eq.~\eqref{aaeqfk104} becomes a Gaussian distribution for any observation time $t$, indicating normal diffusion. However, in the presence of any disturbance, we observe super-diffusion in the long-time limit.
\item According to Eq.~\eqref{aaeqfk104} or \eqref{aaeqfk102}, we get the asymptotic behavior of the mean of the position
$\langle x(t)\rangle\sim at/\langle\tau\rangle,$
growing linearly with time $t$.
As expected, Eq.~\eqref{aaeqfk104} gives an infinite MSD, which is certainly not possible. In Ref.~\cite{Wang2019Transport} we find that the MSD is sensitive to rare events, while Eq.~\eqref{aaeqfk104} deals with typical fluctuations describing the central part of the positional distribution. A detailed discussion about the mean squared displacement (MSD) and fractional moments will be given in Appendix \ref{18ctrwSect7}.

\item According to Eqs.~\eqref{aaeqfk102} and \eqref{wqee1021}, we have
\begin{equation}\label{eq03}
\mathcal{P}(k,t)= \exp\left[ - D k^2 t -i k V t + S (-i k)^\alpha t\right].
\end{equation}
Instead of considering $\mathcal{P}(x,t)$, we investigate
\begin{equation}
\mathcal{Q}(k,t)=\exp(i k V t )\mathcal{P}(k,t) = \exp\left[ -  D k^2 t + S (-i k)^\alpha t\right].
\label{eq04}
\end{equation}
Performing the Laplace transform with respect to $t$, we find
\begin{equation}
\mathcal{Q}(k,s)=\mathcal{P}\left(k,s-iVk\right)=\frac{1}{s+\frac{\sigma^2}{2\langle\tau\rangle}k^2-\frac{ a^\alpha}{\bar{t}}(-ik)^\alpha}.
\label{QKSFL}
\end{equation}
The last term of Eq.~\eqref{QKSFL} demonstrates the  competition between normal diffusion and sup-diffusion. It can be seen that when $\sigma^2k^2/(2\langle\tau\rangle)\gg |(-ik)^\alpha| a^\alpha/\bar{t}$, i.e., $|k|\gg (a^\alpha 2\langle\tau\rangle/(\bar{t}\sigma^2))^{1/(2-\alpha)}$, as mentioned we get the Gaussian distribution. For the opposite limit, the L{\'e}vy stable law  is derived.

\item  The L{\'e}vy stable law and Gaussian distribution, which are the limit laws for solving Eq.~\eqref{aaeqfk104}, are controlled by the dimensionless parameter $\overline{\sigma}^2$. More details on this can be found in Sec.~\ref{18ctrwsect2} in terms of CTRW.
%

\item We further investigate the situation when the bias is negative, i.e., $a<0$. Substituting Eq.~\eqref{wqee1021} into Eq.~\eqref{eq03},
we have
\begin{equation*}\label{eq101}\widetilde{\mathcal{P}}(k,t)=\exp\left(-\frac{\sigma^2}{2\langle\tau\rangle}k^2t-ik\frac{a}{\langle\tau\rangle}t+\frac{a^\alpha}{\bar{t}}(-ik)^\alpha t\right).
\end{equation*}
Arranging  the above equation, we find
\begin{equation}\label{eq102}
\widetilde{\mathcal{P}}(k,t)=\exp\left(-\frac{\sigma^2 k^2t}{2\langle\tau\rangle}-ik\frac{at}{\langle\tau\rangle}+\frac{(-a)^\alpha}{\bar{t}}(ik)^\alpha t\right).
\end{equation}
Taking derivative with respect to $t$ and then performing the inverse Fourier transform, Eq.~\eqref{eq102} leads to
\begin{equation}\label{eq105}
\frac{\partial }{\partial t}\mathcal{P}=\frac{\sigma^2}{2\langle\tau\rangle}\frac{\partial^2 }{\partial x^2}\mathcal{P}-\frac{a}{\langle\tau\rangle}\frac{\partial }{\partial x}\mathcal{P}+\frac{(-a)^\alpha}{\bar{t}} \frac{\partial^\alpha }{\partial x^\alpha}\mathcal{P}.
\end{equation}
In the case of a positive bias, we have the right Riemann-Liouville derivative in Eq.~\eqref{aaeqfk104}. While, if the bias is negative, the fractional operator in Eq.~\eqref{aaeqfk104} is replaced by the left Riemann-Liouville derivative [see Eq.~\eqref{Weyl2}], controlling the right fat tail of the positional distribution.


\item When waiting times have an infinite mean, i.e., Eq.~\eqref{ldeq3201111} with $0<\alpha<1$, and the displacement is still drawn from Gaussian distribution Eq.~\eqref{18eq104}, the corresponding fractional advection diffusion equation \cite{Metzler1999Anomalous,Metzler1999Deriving,Metzler2000random,Barkai2001Fractional,Vladimir2013Fractional} is
\begin{equation}\label{eeeesssa}
\frac{\partial}{\partial t}\mathcal{P}(x,t)=~\!_0\mathcal{D}_t^{1-\alpha}\Big[D\frac{\partial^2}{\partial x^2}-V\frac{\partial}{\partial x}\Big]\mathcal{P}(x,t),
\end{equation}
where $D=\sigma^2/b_\alpha$ and $V=a/b_\alpha$ with $b_\alpha=\tau_0^\alpha\Gamma(1-\alpha)$. In Eq.~\eqref{eeeesssa}, the fractional operator $\!_0\mathcal{D}_t^{1-\alpha}$ is in reference to time rather than the space. The fractional time operator $_0\mathcal{D}_t^{1-\alpha}$ called the  Riemann-Liouville time derivative is as follows \cite{Podlubny1999Fractional,Deng2020Modeling}
\begin{equation}
_0\mathcal{D}_t^{1-\alpha} g(t)=\frac{1}{\Gamma(\alpha)}\frac{\partial}{\partial t}\int_{0}^{t}\frac{g(\tau)}{(t-\tau)^{1-\alpha}}d\tau
\end{equation}
with $0<\alpha<1$.
Further, when $0<\alpha<1$, the positional distribution has a right fat tail. While, in the context of $1<\alpha<2$, the positional distribution has a left one.
It can be seen that the forms of fractional equations with different $\alpha$ show great differences.

\item Subordination discussed here is vastly different from the case for $0<\alpha<1$. When $1<\alpha<2$, the number of renewals follows the two-sided L\'evy stable law with an infinite variance, instead of the one-sided L\'evy distribution (also called the Mittag-Leffler distribution).

\item  Solutions we discussed are valid for free boundary conditions. One can use the subordination scheme Eq.~\eqref{18eq502} to consider other cases. The key idea is that the number of steps in the long time limit is distributed according to the L\'evy law given in Eq.~\eqref{18eq501a}. Then a normal process is considered,  for example  diffusion in a finite medium, with the operational time $N$. The L\'evy central limit theorem is then used to transform  the operational time $N$ (also the number of steps) to the laboratory time $t$.

\item Benson et al. proposed an equation called fractional advection dispersion  equation to describe  contaminant source \cite{Benson2001Fractional,Yong2016Backward}
\begin{equation}\label{eqzhang2016}
\frac{\partial }{\partial t}\mathcal{P}=-V\frac{\partial }{\partial x} \mathcal{P}+K\left[p \frac{\partial^{\alpha} }{\partial x^{\alpha}}\mathcal{P}+q\frac{\partial^{\alpha} }{\partial (-x)^{\alpha}}\mathcal{P}\right],
\end{equation}
which is obtained from L{\'e}vy flights; see Fig.~\ref{bigjump} (a). If $D=0$ and $p=0$, Eq.~\eqref{aaeqfk104}  is consistent with Eq.~\eqref{eqzhang2016}.
In Ref. \cite{zhang2005mass}, the authors use $V=0.8$m/h, $D=0$, $K=2.8{\rm m}^{1.51}/{\rm h}$, $q=1$, and $p=0$  to fit experimental contaminant data in terms of the breakthrough curves. According to this, one may interpret the data as coming from L{\'e}vy flights. On the other hand, here we want to stress that these data may stem from CTRW model with a strong bias in the long time limit, i.e., these experimental data are consistent with CTRW with a fat-tailed waiting time distribution. In that sense, it indicates that both L{\'e}vy flights and CTRW can be used as a tool to describe the contaminant in disorder systems. This holds as long as we have information on the spreading packets and no physical insight into the underlying  trajectories.
\end{enumerate}

\section{Applications}\label{applications}
As mentioned before, fractional diffusion equations are widely used as a tool to describe the dispersion of contaminants and other phenomena observed in non-equilibrium systems.
For that, we explore several applications of FADAE \eqref{aaeqfk104}. These applications encompass
calculating FADAE in two dimensions, determining the positional distribution with the time-dependent bias and variance, analyzing breakthrough curves \cite{Alon2017Time,Metzler2022Modelling}, estimating the first passage time, and comparing them to CTRW dynamics.
%
%
%
%
%

\subsection{FADAE in two dimensions}\label{18ctrwSecttwoD101}
\begin{figure}[htb]
  \centering
  \includegraphics[width=0.5\textwidth]{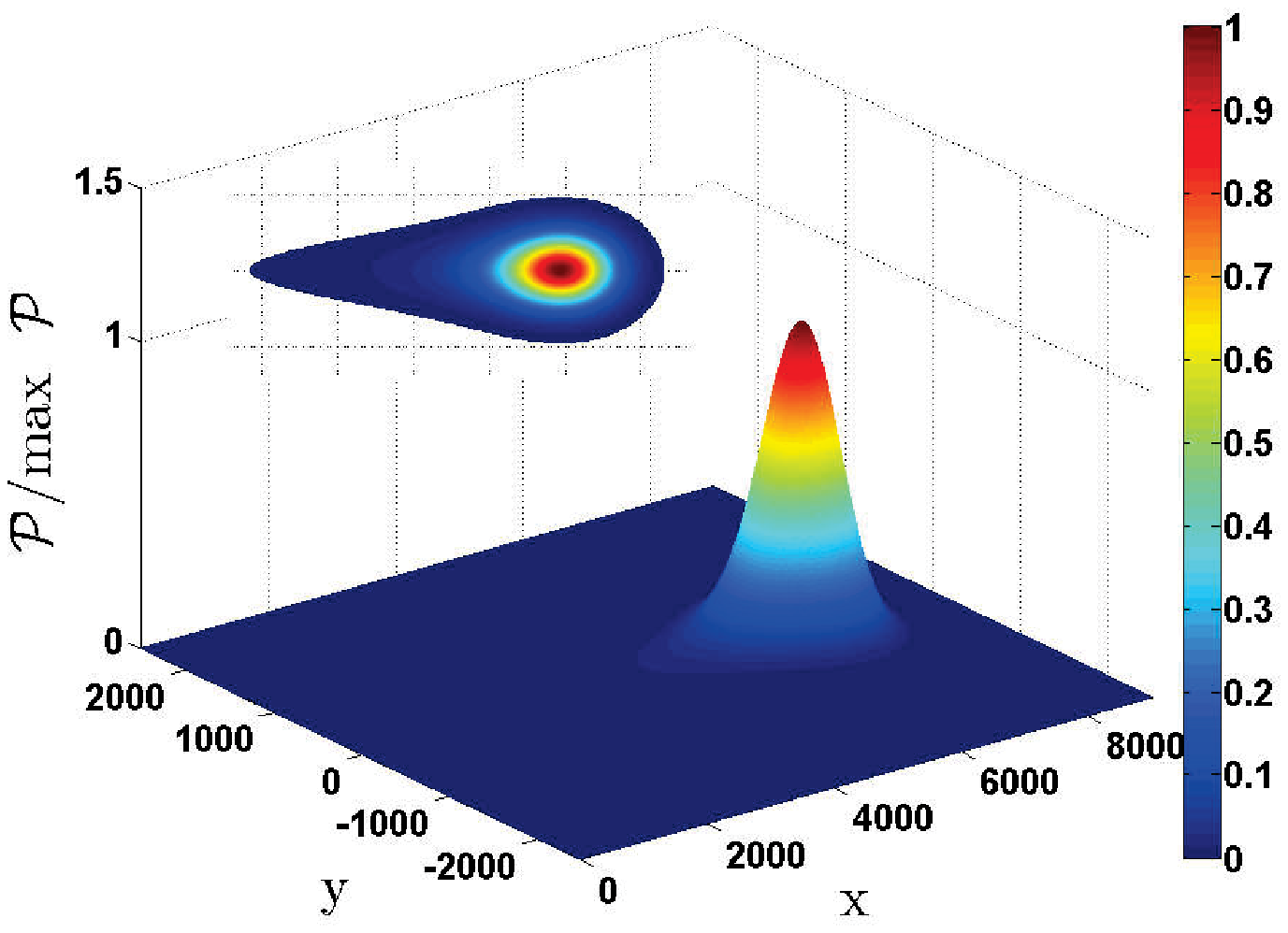}\\
  \caption{The dynamic of $\mathcal{P}(x,y,t)/{\rm max}\mathcal{P}(x,y,t)$ obtained from Eq.~\eqref{18ctrwSecttwoD106} for the fractional advection-diffusion process in two dimensions. However, in the inset, the results are projected. The parameters of CTRW are $\tau_0=0.1$, $\alpha=1.5$, $a_x=2$, $t=1000$ and $\sigma_x=\sigma_y=5$. }
\label{pxt3dnew}
\end{figure}

\begin{figure}[htb]
  \centering
  \includegraphics[width=0.5\textwidth]{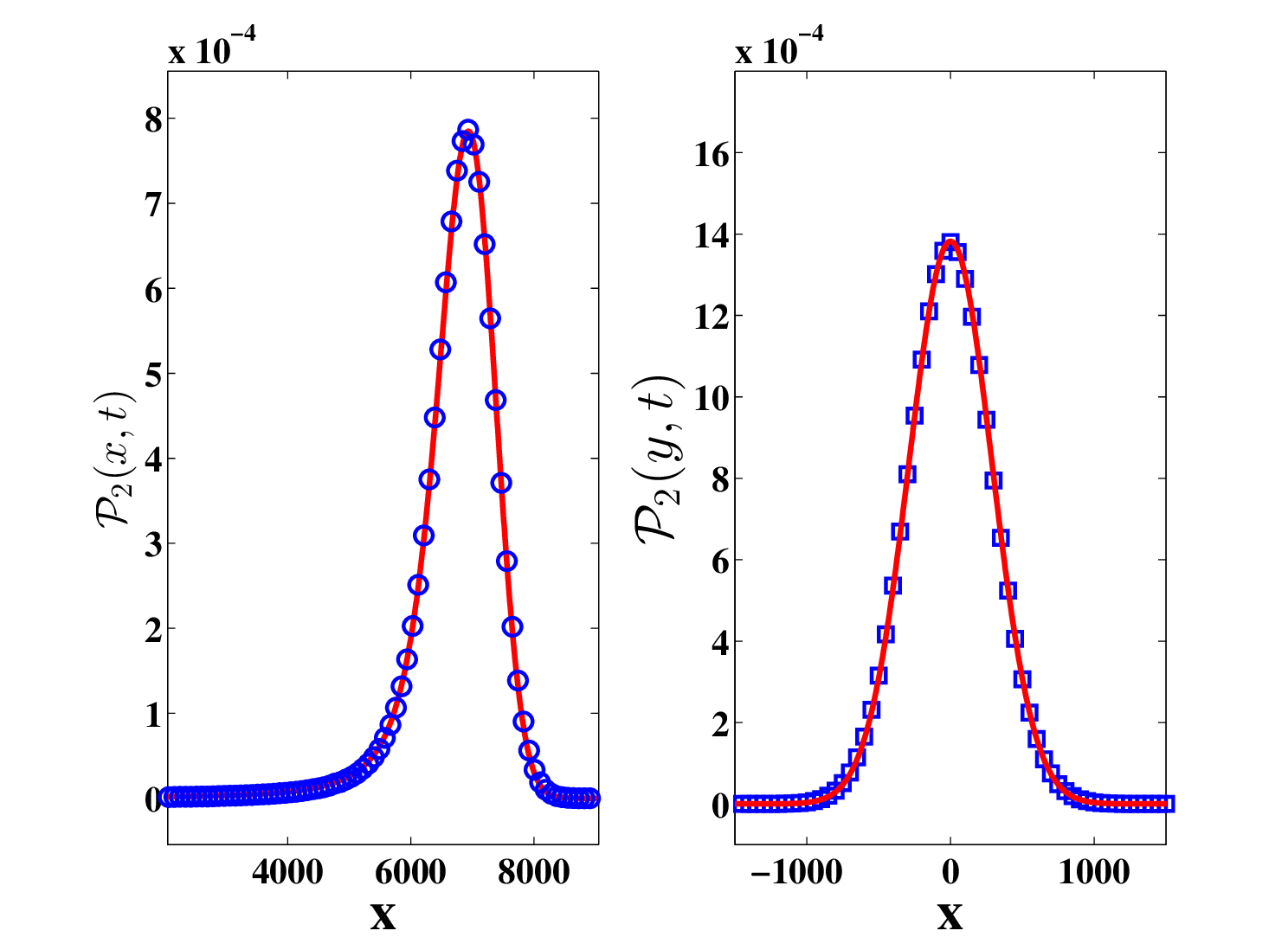}\\
  \caption{Plot of marginal distribution Eq.~\eqref{18ctrwSecttwoD106} compared with CTRW simulations. Here we denote $\mathcal{P}_2(x,t)$ and $\mathcal{P}_2(y,t)$ as the marginal distribution of $\mathcal{P}(x,y,t)$ with respect to $x$ and $y$, respectively. It can be seen that $\mathcal{P}_2(x,t)$ is the same as the solution of Eq.~\eqref{aaeqfk104} and $\mathcal{P}_2(y,t)$ is a Gaussian distribution.
The parameters are the same as in Fig.~\ref{2DPXTFromXY}.}
\label{2DPXTFromXY}
\end{figure}

Here we consider the two dimensional FSDAE which  is  not only of theoretical significance, but also  potential application value \cite{Berkowitz2001Application,Berkowitz2006Modeling,Berkowitz2009Exploring,Alon2017Time}. For simplification, the bias of the system is only in $x$ direction and the equation reads
\begin{equation}\label{18ctrwSecttwoD106}
\setlength{\parindent}{10em} \frac{\partial}{\partial t}\mathcal{P}(x,y,t)=D_x\frac{\partial^2}{\partial x^2}\mathcal{P}(x,y,t)-V_x\frac{\partial}{\partial x}\mathcal{P}(x,y,t)+S_{\smallskip x}\frac{\partial^\alpha}{\partial (-x)^\alpha} \mathcal{P}(x,y,t)+D_y\frac{\partial^2}{\partial y^2}\mathcal{P}(x,y,t).
\end{equation}
Here $D_x$, $V_x$, $S_x$, and $D_y$ are constants. Taking double Fourier transforms, $x\to k_x$ and $y\to k_y$,  we get
\begin{equation}\label{18ctrwSecttwoD105}
 \widetilde{\mathcal{P}}(k_x,k_y,t)=\exp\left(-D_x(k_x)^2t-iV_xk_xt+(-ik_x)^\alpha tS_x-D_y(k_y)^2t\right).
\end{equation}
Using convolution properties of Fourier transform, Eq.~\eqref{18ctrwSecttwoD105} yields
\begin{equation}\label{18ctrwSecttwoD104}
 \mathcal{P}(x,y,t)=\int_{-\infty}^\infty \mathcal{L}_{\alpha}(z)\frac{\exp\left(-\frac{(x-V_xt-(tS_x)^{1/\alpha}z)^2}{4D_xt}\right)}{\sqrt{4\pi D_xt}}\frac{\exp\left(-\frac{y^2}{4D_y t}\right)}{\sqrt{4\pi D_yt}}dz,
\end{equation}
where $\mathcal{L}_{\alpha}(z)$ is the non-symmetric L{\'e}vy stable distribution Eq.~\eqref{deflevystable}.
Below we obtain four transport coefficients given in Eq.~\eqref{18ctrwSecttwoD106}.
In the language of the CTRW model,  the displacement follows
\begin{equation}\label{18ctrwSecttwoD102}
  f(x,y)=\frac{1}{\sqrt{2(\sigma_x)^2\pi}}\exp\left(-\frac{(x-a_x)^2}{2(\sigma_x)^2}\right)\times \frac{1}{\sqrt{2(\sigma_y)^2\pi}}\exp\left(-\frac{y^2}{2(\sigma_y)^2}\right),
\end{equation}
where $a_x$, $\sigma_x\neq 0$, and $\sigma_y\neq 0$ are constants.
In double Fourier spaces, $x\to k_x$ and $y\to k_y$, we get
\begin{equation}\label{18ctrwSecttwoD103}
  \widetilde{f}(k_x,k_y)=\exp\left(-ik_xa_x-\frac{1}{2}(\sigma_x)^2(k_x)^2-\frac{1}{2}(\sigma_y)^2(k_y)^2\right).
\end{equation}
For the waiting time, we continue to utilize the fat-tailed power law distribution Eq. \eqref{ldeq3201111}. Using Eq. \eqref{wqee1021} and the same arguments as before, we find
\begin{equation}\label{wqee1021addwq}
 D_x=\frac{\sigma_x^2}{2\langle\tau\rangle},~~~
 V_x=\frac{a_x}{\langle\tau\rangle},~~~
 S_x=\frac{(a_x)^\alpha}{\overline{t}},~~~D_y=\frac{\sigma_y^2}{2\langle\tau\rangle}.
\end{equation}
It can be seen that when the external force is only in the $x$ direction, the fat tail of the packet of spreading particle $\mathcal{P}(x,y,t)$ is with respect to $x$. Note that the exact solution of Eq.~\eqref{18ctrwSecttwoD106} is Eq.~\eqref{18ctrwSecttwoD104}, i.e., $P_{\rm{CTRW}}(x,y,t)$ in the long time limit. What we want to stress is that when the bias is also in  the $y$ direction, the fractional space operator with respect to $y$ should be added.
See the solution of Eq.~\eqref{18ctrwSecttwoD106} in Fig.~\ref{pxt3dnew} and the marginal distribution in Fig.~\ref{2DPXTFromXY}.
%


\subsection{Propagator with the time-dependent bias and variance}
\begin{figure}[h]
  \centering
  \includegraphics[width=0.5\textwidth]{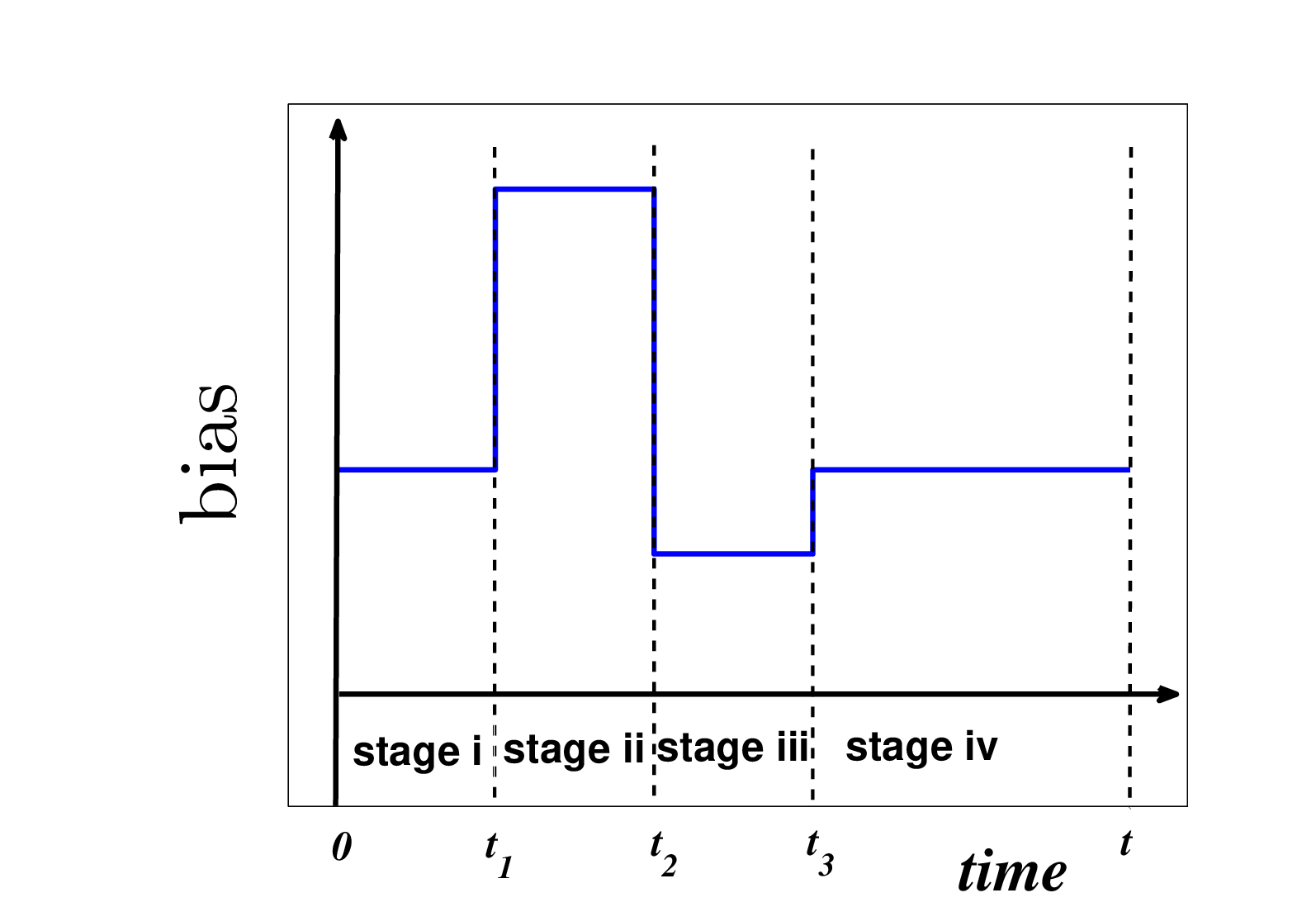}\\
  \caption{Illustrations of the time-dependent bias, e.g., $a(t)$, when the process
begins at time $t=0$ and ends at time $t$. In our simulations, the mean displacements of four states, i.e., bias, are determined by $\{a,4a\mu/(\mu+1),a/(2/3+\mu),a\}$. }
\label{illustrationOfBias}
\end{figure}
Motivated by \cite{Alon2017Time}, we consider the time-dependent bias and variance determined by four states. In some sense, assumptions are more closely aligned with the real-world spread of pollutants.
This idea yields some interesting results, for example complex structure of breakthrough curves; see below.
We suppose that the rapid injection of particles is done immediately after starting observing the process. In other words,  the  initial positions of all particles are $\mathcal{P}(x,0)=\delta(x)$.
Here we simulate the particles consisting of four states: (i) after the injection of the particles at time $t=0$, they undergo CTRW processes with a constant bias determined by $a$ in time interval $(0,t_1)$ (here $a$ is the mean of displacements), (ii) during the time interval $t_1<t< t_2$ we increase the force, namely replace the bias $a$ with $4a \mu/(\mu+1)$ for $\mu\geq1/3$, (iii) we further decrease the bias to $a/(2/3+\mu)$ from time $t_2$ to $t_3$, and (iv) finally we return to the state (i) from time $t_3$; see Fig.~\ref{illustrationOfBias} for the statistics of the bias versus time. Here $\mu$ is a  constant that controls the strength of the force.
Let $\sigma=5$ for states (i), (ii), and (iv), i.e., $\sigma_1=\sigma_2=\sigma_4=5$. While, for state (iii), we use $\sigma_3=0.1$.
In particular, when $\mu=1/3$, the bias  of all states mentioned above is the same but the variance is still time-dependent. Note that the elapsed time of each state should be a bit long, otherwise one can find that the difference between them is not very large. In Fourier space, the initial condition satisfies $\widetilde{\mathcal{P}}(k,0)=1$.
In view of the special expression of $\widetilde{\mathcal{P}}(k,t)$,  $\widetilde{\mathcal{P}}(k,t)$ for different states is as follows
\begin{equation}\label{aaeqfk301}
\widetilde{\mathcal{P}}(k,t)=\left\{
  \begin{array}{ll}
    \exp(-c_{11}k^2-ic_{12}k+c_{13}(-ik)^\alpha), & \hbox{$0<t\leq t_1$;} \\
   \exp(-c_{21}k^2-ic_{22}k+c_{22}(-ik)^\alpha), & \hbox{$t_1<t\leq t_2$;} \\
   \exp(-c_{31}k^2-ic_{32}k+c_{33}(-ik)^\alpha)  , & \hbox{$t_2<t\leq t_3$;} \\
   \exp(-c_{41}k^2-ic_{42}k+c_{43}(-ik)^\alpha) , & \hbox{$t_3<t$,}
  \end{array}
\right.
\end{equation}
with

\begin{equation}\label{aaeqfk302}
c_{m1}=\left\{
  \begin{array}{ll}
   t\frac{\sigma_1^2}{2\langle\tau\rangle} , & \hbox{$m=1$;} \\
   t\frac{\sigma_2^2}{2\langle\tau\rangle}+\frac{t_1(\sigma_1^2-\sigma_2^2)}{2\langle\tau\rangle}, & \hbox{$m=2$;} \\
   t\frac{\sigma_3^2}{2\langle\tau\rangle}+\frac{t_1(\sigma_1^2-\sigma_2^2)}{2\langle\tau\rangle}+\frac{t_2(\sigma_2^2-\sigma_3^2)}{2\langle\tau\rangle}, & \hbox{$m=3$;} \\
   t\frac{\sigma_4^2}{2\langle\tau\rangle}+\frac{t_1(\sigma_1^2-\sigma_2^2)}{2\langle\tau\rangle}+\frac{t_2(\sigma_2^2-\sigma_3^2)}{2\langle\tau\rangle}+\frac{t_3(\sigma_3^2-\sigma_4^2)}{2\langle\tau\rangle}, & \hbox{$m=4$,}
  \end{array}
\right.
\end{equation}

\begin{equation}\label{aaeqfk303}
c_{m2}=\left\{
  \begin{array}{ll}
   a_1\frac{t}{\langle\tau\rangle}, & \hbox{$m=1$;} \\
   \left[a_1t_1+a_2(t-t_1)\right]\frac{1}{\langle\tau\rangle}, & \hbox{$m=2$;} \\
  \left[a_1t_1+a_2(t_2-t_1)+a_3(t-t_2)\right]\frac{1}{\langle\tau\rangle}, & \hbox{$m=3$;} \\
  \left[a_1t_1+a_2(t_2-t_1)+a_3(t_3-t_2)+a_4(t-t_3)\right]\frac{1}{\langle\tau\rangle} , & \hbox{$m=4$,}
  \end{array}
\right.
\end{equation}
and
\begin{equation}\label{aaeqfk304}
c_{m3}=\left\{
  \begin{array}{ll}
   a_1^\alpha t\frac{1}{\overline{t}}, & \hbox{$m=1$;} \\
   \left[a_1^\alpha t_1+a_2^\alpha(t-t_1)\right]\frac{1}{\overline{t}}, & \hbox{$m=2$;} \\
  \left[a_1^\alpha t+a_2^\alpha(t_2-t_1)+a_3^\alpha(t-t_2)\right] \frac{1}{\overline{t}}, & \hbox{$m=3$;} \\
 \left[a_1^\alpha t_1+a_2^\alpha(t_2-t_1)+a_3^\alpha(t_3-t_2)+a_4^\alpha(t-t_3)\right] \frac{1}{\overline{t}}, & \hbox{$m=4$.}
  \end{array}
\right.
\end{equation}
Here $m=1,2,3,4$ is related to the number of states.
Recall that $\bar{t} =\langle\tau\rangle^{1 +\alpha}/[(\tau_0)^\alpha|\Gamma(1 - \alpha)|]$. In  calculations, the main idea is that the final position of each state is treated as the initial position of the next  stage.
The  inverse Fourier transform of Eq.~\eqref{aaeqfk301} yields
\begin{equation}\label{aaeqfk305}
\mathcal{P}(x,t)=\int_{-\infty}^{\infty}\frac{1}{\sqrt{4\pi c_{m1}}}\exp\left[-\frac{(x-y-c_{m2})^2}{4c_{m1}}\right]\frac{1}{(c_{m3})^{1/\alpha}}\mathcal{L}_{\alpha}\left[\frac{y}{(c_{m3})^{1/\alpha}}\right]dy,
\end{equation}
describing the positional distribution of the mentioned four states.
Note that when $0<t<t_1$, the solution Eq.~\eqref{aaeqfk305} is the same as  Eq.~\eqref{aaeqfk101} or that of Eq.~\eqref{aaeqfk104}. Below, we  use Eq.~\eqref{aaeqfk305} to predict breakthrough curves \cite{Cortis2004Anomalous}.

\subsection{Breakthrough curves }

\begin{figure}[htp]
\begin{center}
    \includegraphics[width=0.5\textwidth]{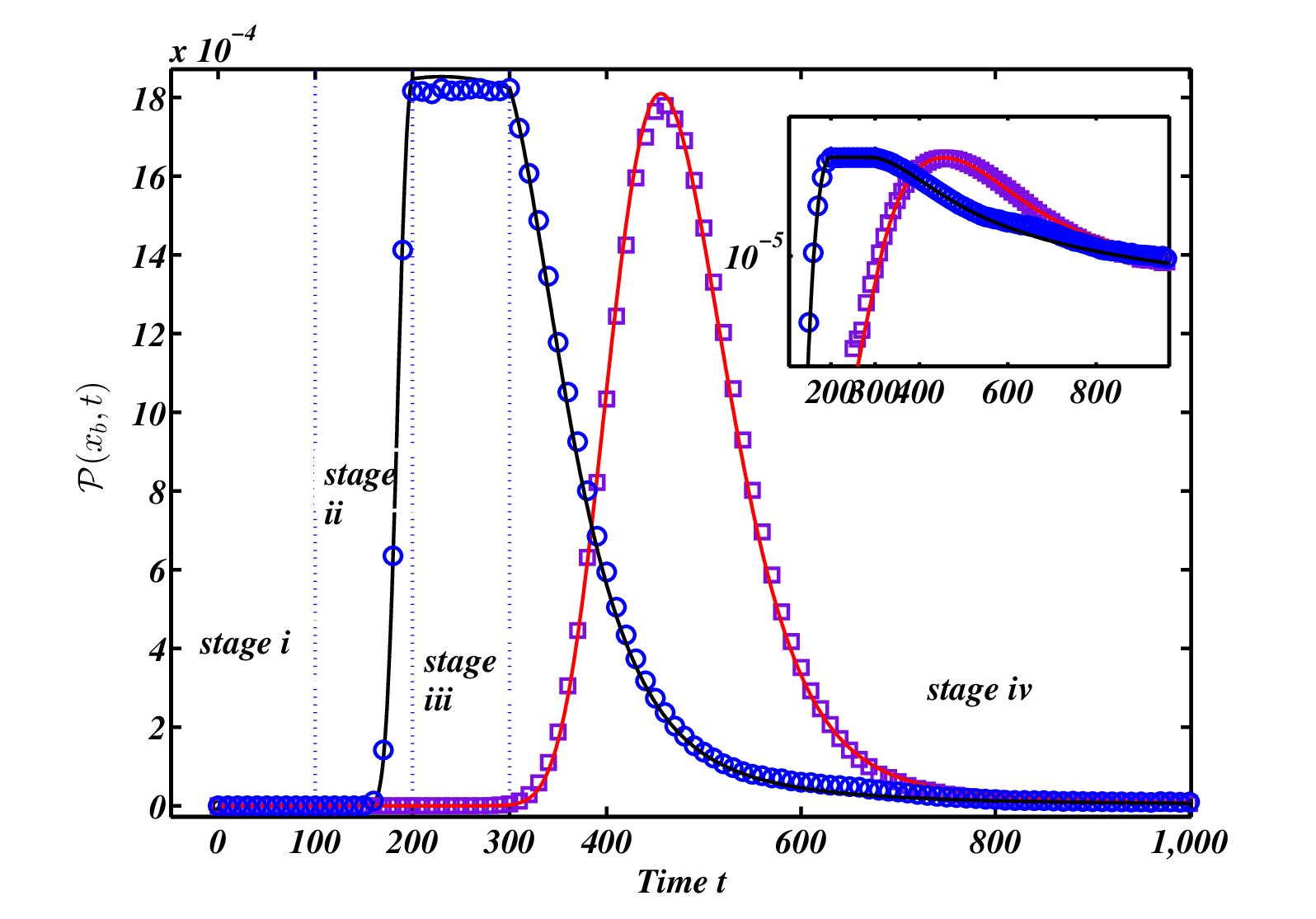}
    \caption{Theoretical predictions of breakthrough curves together with the simulations plotted by the symbols. The solid lines are the theories based on Eq.~\eqref{aaeqfk305}. Time intervals of the four states are $100, 100, 100, 700$, respectively. Packets of particles are  asymmetric in the long time limit, which are the same for both constant and time-dependent cases. Semi-log scales are represented to show asymmetric properties and the heavy tail; see the inset. Breakthrough curves are observed at the position $x_b=1600$.
 Here the parameters are $\mu=1/3$ (red line), $\mu=6$ (black line), $\tau_0=0.1, \alpha=1.5$, $\sigma_1=\sigma_2=\sigma_4=5$, and $\sigma_3=0.1$.}
\label{FixXwithsi5}
\end{center}
\end{figure}

Contamination spreading, as one of the most crucial  problems ranging from environment to agriculture, has attracted a lot of attention \cite{Berkowitz2006Modeling,Berkowitz2009Exploring,Alon2017Time,Generalized2022Ricardo,Doerries2022Rate}. In real systems, how to quantity the contaminant transport is quite a challenge for  researchers  since  the processes are so complex. In \cite{Alon2017Time}, Nissan, Dror and Berkowitz considered the spreading particle with changing conditions to illustrate the positive and negative effects of the ambient environment. It is no wonder that the changing force fields are quite common and vital in the real world.
Theoretical predictions about breakthrough curves, which can be directly measured in experiments, were made using the CTRW particle tracking approach in \cite{Alon2017Time}.  While, here we choose Eq.~\eqref{aaeqfk305} as a tool to describe breakthrough curves. The breakthrough curves are measured in the sense of the distribution, indicating the probability of particles being at $x_b$ at a specific time $t$.
The differences and similarities between the constant and the time-dependent force are quite interesting. When particles experience  small disturbances or the total time of the first three states is short, the response of time-dependent force, calculated by the density of the position, is weak except for a `shift'.
However, when we increase the total time of the first three states, the mentioned two cases exhibit great differences; see Fig.~\ref{FixXwithsi5}. Note that in our discussions the detection point is $x_b=1600$. Choosing the suitable detection point, i.e., $x_b$, is a critical factor in observing the distinctive structure of the particle packet. For a short time $t$, such as $t=10$, it can be seen that $\mathcal{P}(x_b,t)\to 0$ since all the particles are on the way to the site $x_b$ pushed by the force and need more time to visit the position $x_b$. As particles enter the second state, they are moving faster than the state (i) for $\mu=6$, resulting in a rapid increase in the probability of the position. During the time interval $(t_2,t_3)$, the disturbance and the variance are weak, leading to a more or less flat breakthrough curve. At the end of this state, there are still numerous particles being located at the left side of $x_b$. In the last state, $\mathcal{P}(x_b,t)$ slowly  tends to zero.

%

\subsection{First passage time}
So far we considered problems with free boundary conditions. The boundary conditions for fractional space diffusion equations are generally non-trivial. This is because fractional operators are non-local in space. Our approach to the problem is to use the subordination method, namely rely on Eqs.~(\ref{18eq500},\ref{18eq501a}).
The idea is as follows: We showed already that fluctuations of the number of renewals $N$ is given by L\'evy statistics. We may think about the whole
problem as a normal process, where $N$ is an operational time. Then this normal process is transformed from $N$ to the laboratory time $t$. By normal process, we mean, for example, Brownian diffusion with advection in a finite domain, with  reflecting or absorbing boundary conditions.

 Here we consider as an example the case of the first passage time  problem \cite{Redner2001guide,Metzler2012Levy,Wardak2020First,Zan2021First,Holl2022Controls}. The first passage time serves as a significant tool for describing the time required by a particle to reach a specific point or an absorbing barrier.
For standard diffusion this means that we use an absorbing boundary condition (see below). We will later show how to use subordination to obtain the solution for the anomalous process.

 Before doing so we will briefly delve into the method of images. It was shown previously, that for L{\'e}vy flights the image method fails \cite{Chechkin2003First,Sokolov2004Non,Palyulin2019First,Padash2019First}. However, in our case, we do not have any L{\'e}vy flight as an underlying process. Instead, we use fat-tailed distributions for waiting times. We will first show that the method of images generally fails in our case, however when the bias is weak it is a reasonable approximation. We will then turn to the more general solution using subordination.

\subsubsection{Recap for the image method and it's failure}

Let us determine the first passage probability based on Eq.~\eqref{aaeqfk104} for a diffusing particle that starts at $x_0=0$. As mentioned, we use the image method and show that while it is generally not a good approximation, it works well when the bias is small.
Here we assume that the absorbing boundary condition is $x_{ab}$ and use $C(x,t)$  to denote the concentration. Thus, when $x\geq x_{ab}$,  we have $C(x,t)=0$. The most appealing approach, dealing with the absorbing boundary condition, is the method of images, which stems from electrostatics. As mentioned we  will soon show, following Eq.~\eqref{aaeqfk104}, that this method does not work when the bias is not terribly small.
In real space, the image method leads to \cite{Barkai2001Fractional,Redner2001guide},
\begin{equation}\label{FPT1001}
C(x,t)=\mathcal{P}(x,t)-W\mathcal{P}(x-2x_{ab},t),~-\infty<x\leq x_{ab},
\end{equation}
where the unknown parameter $W$ is defined below.
Here $\mathcal{P}(x,t)$ denotes the positional distribution without absorbing condition, for example, below we use the solution of Eq.~\eqref{aaeqfk104} to check the validity of the image method. It can be seen that the concentration is the difference between $\mathcal{P}(x,t)$ and $\mathcal{P}(x-2x_{ab},t)$ with a weight $W$ determined by $x_{ab}$. Recall that $C(x,t)$ vanish on $x_{ab}$, i.e., $C(x,t)=0$. Then $W$ obeys
$$\mathcal{P}(x_{ab},t)-W\mathcal{P}(-x_{ab},t)=0.$$
Thus, we get with the image method
\begin{equation}\label{FPT1002}
W=\frac{\mathcal{P}(x_{ab},t)}{\mathcal{P}(-x_{ab},t)}.
\end{equation}
Consider the well-studied case of a particle undergoing normal diffusion (Eq.~\eqref{aaeqfk104} with $S=0$) with an absorbing condition at $x=x_{ab}$. Thus, $\mathcal{P}(x,t)$ in the absence of the absorbing boundary condition follows Gaussian distribution
\begin{equation}
\mathcal{P}(x,t)=\frac{1}{\sqrt[]{4\pi D t}}\exp\Big(-\frac{(x-Vt)^2}{4D t}\Big),
\end{equation}
which leads to
\begin{equation}
W=\exp\left(x_{ab}\frac{V}{D}\right)
\end{equation}
according to Eq.~\eqref{FPT1002}.
Note that in the general case $W$ is determined by $x_{ab}$, $D$, $V$ and $S$. 

Based on Eqs.~\eqref{FPT1001} and \eqref{FPT1002}, mathematically, the concentration is as follows
\begin{equation}\label{FPT1004}
C(x,t)=\mathcal{P}(x,t)-\frac{\mathcal{P}(x_{ab},t)}{\mathcal{P}(-x_{ab},t)}\mathcal{P}(x-2x_{ab},t)
\end{equation}
with $-\infty<x\leq x_{ab}$. Equation \eqref{FPT1004} is demonstrated in Fig.~\ref{ObPXT} in Appendix \ref{CTRWAPPENc} for different absorbing conditions.
 We see here that for weak fields the image theory works  since the process is almost Gaussian.

\subsubsection{Subordinating the first passage problem}

\begin{figure}[htb]
  \centering
  \includegraphics[width=9cm, height=6cm]{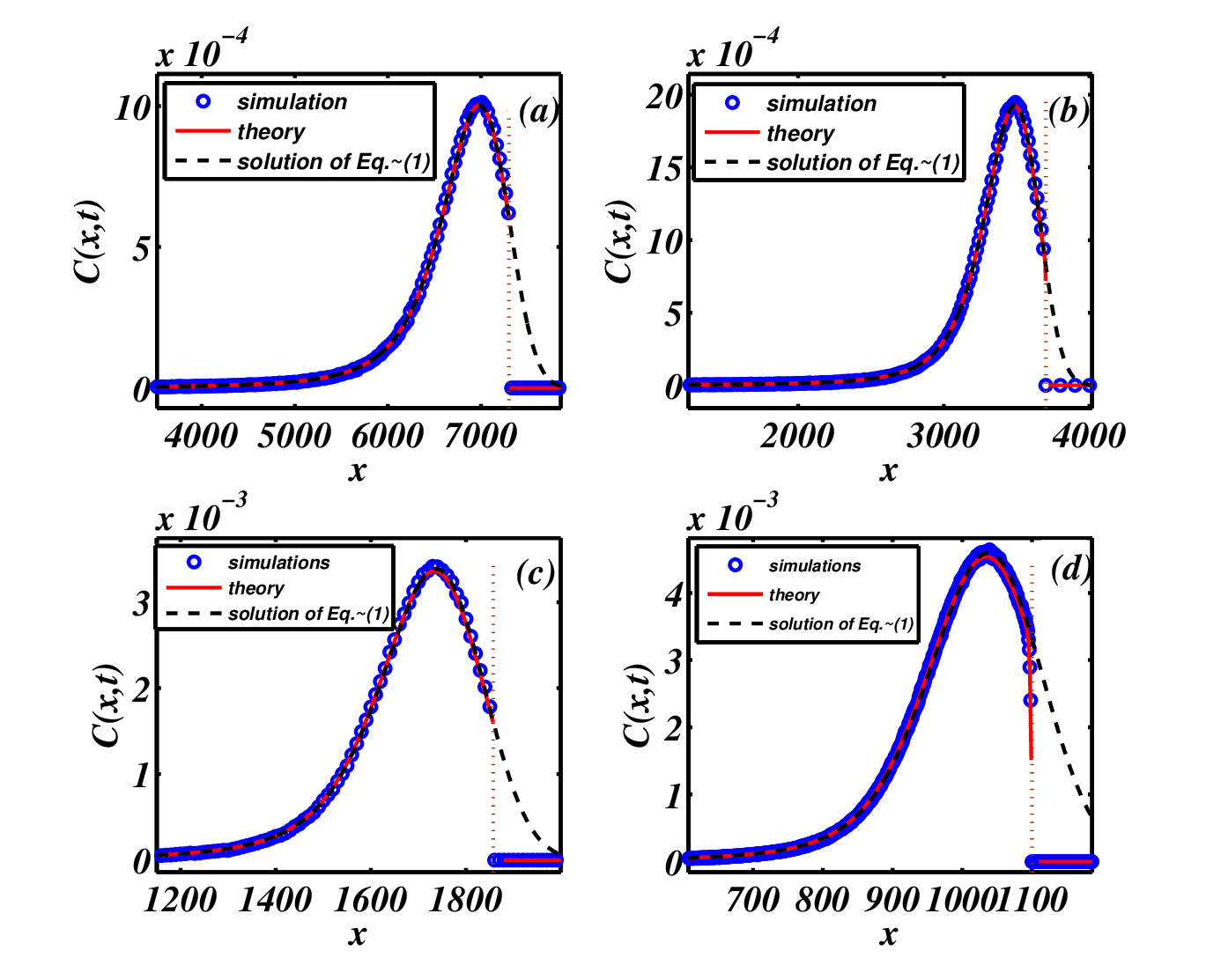}\\
  \caption{Plot of concentration $C(x,t)$ using Eq.~\eqref{18eq502add} for different biases and locations of absorbing boundary conditions. Here we choose $t=1000$, $\tau_0=0.1$, $\alpha=3/2$, and $10^7$ realizations. We use $a=2, x_{ab}=7300$; $a=1, x_{ab}=3700$; $a=1/2, x_{ab}=1850$; and $a=3/10, x_{ab}=1100$ for subplot (a) to (d), respectively. The Solution of Eq.~\eqref{aaeqfk104} without absorbing condition is plotted by the dashed lines. }
\label{InfiniteSumImage}
\end{figure}

To solve this problem, we use a subordination technique that involves utilizing the method of images on the $\chi_N(x)$.
Based on Eq.~\eqref{18eq500}, the discrete form of $C(x,t)$ follows
\begin{equation}\label{FPT1006}
C(x,t)\sim \sum_{N=0}^\infty Q_t(N) \chi^{*}_N(x),
\end{equation}
describing the subordinated process $x$ as a function of time $N$.
Here $Q_t(N)$ is the PDF of the number of the renewals given in Eq.~\eqref{18eq501a} and $\chi^{*}_N(x)$ is the solution of the ordinary Fokker-Planck equation
\begin{equation}\label{18eq502add1}
\frac{\partial}{\partial N}P(x,N)=a\frac{\partial}{\partial x}P(x,N)-\frac{\delta^2}{2}\frac{\partial^2}{\partial x^2}P(x,N)
\end{equation}
with the initial conditions $P(x,N=0)=\delta(x)$ and different  boundary conditions, be specific, absorbing boundary condition at $x=x_{ab}$. The same approach can be used for other types of boundary conditions. Setting the absorbing condition $x=x_{ab}$ on Eq.~\eqref{18eq502add1} and using the method of images for the solution of Eq.~\eqref{18eq502add1} yield
\begin{equation}\label{FPT1005}
\chi^{*}_N(x)=\frac{\exp\Big(-\frac{(x-aN)^2}{2\sigma^2 N}\Big)}{\sqrt[]{2\pi\sigma^2 N}}
-\exp\left(\frac{2ax_{ab}}{\delta^2}\right)\frac{\exp\Big(-\frac{(x-2x_{ab}-aN)^2}{2\sigma^2 N}\Big)}{\sqrt[]{2\pi\sigma^2 N}}.
\end{equation}
In the long time limit, the continuous form of Eq.~\eqref{FPT1006} follows
\begin{equation}\label{18eq502add}
C(x,t)=\left(\frac{\overline{t}}{t}\right)^{\frac{1}{\alpha}}\int_{0}^{\infty}\mathcal{L}_{\alpha}\left(\frac{N-t/\langle\tau\rangle}{(t/\overline{t})^{1/\alpha}}\right)\chi^{*}_N(x){\rm d}N.
\end{equation}
Equation \eqref{18eq502add} is verified in Fig.~\ref{InfiniteSumImage} showing a perfect match. We now compare solution  Eq. \eqref{18eq502add} and that of Eq. \eqref{aaeqfk104} with free boundary conditions. For that, the solution of Eq.~\eqref{aaeqfk104} without absorbing condition was plotted by dashed black lines. As expected, when $x$ is much smaller than $x_{ab}$, $C(x,t)$ is consistent with Eq.~\eqref{aaeqfk104}, i.e., the random particles are not affected by the absorbing condition, or most of particles have not yet reached the position near $x_{ab}$. In addition, when the bias is strong, Eq.~\eqref{aaeqfk104} agrees with Eq.~\eqref{18eq502add} for $x<x_{ab}$ at least to the naked eye. Roughly speaking, this is because when the bias is strong, no particles are coming back. Thus, the absorbing condition under study loses its role.
While, if $x$ approaches to $x_{ab}$ and the bias is weak, Eq.~\eqref{aaeqfk104} and  Eq.~\eqref{18eq502add} illustrate two different behaviors. See (d) in Fig.~\ref{InfiniteSumImage}.

\begin{figure}[htb]
  \centering
  \includegraphics[width=9cm, height=6cm]{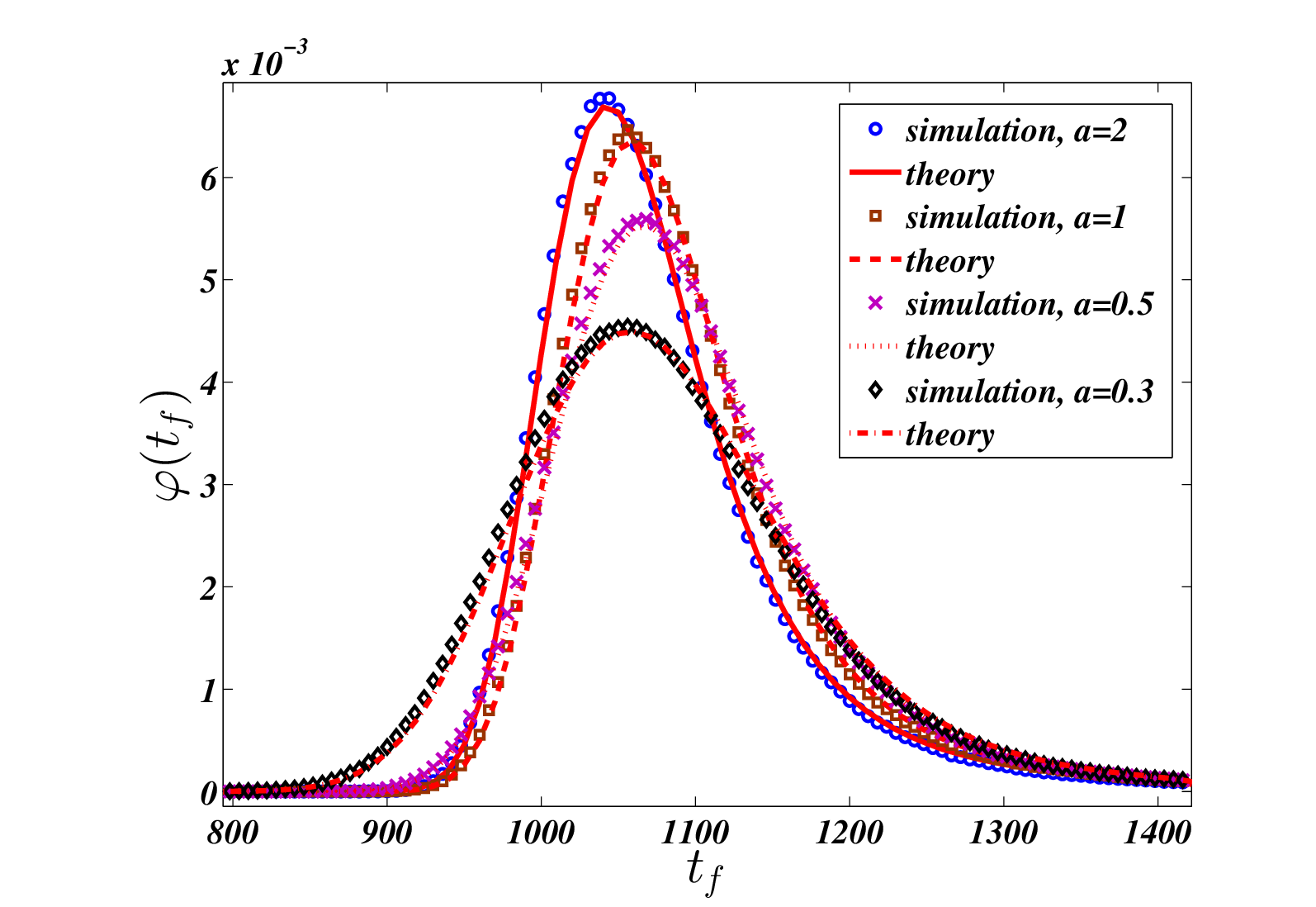}\\
  \caption{Behaviors of the first passage time PDF generated by trajectories of
particles for various $a$. The parameters are the same as in Fig.~\ref{InfiniteSumImage}. The different lines describe the theoretical result
obtained from Eq.~\eqref{FPT100512ad}.}
\label{CTRWFPT}
\end{figure}

%
%
%
%
%
%
%
%
We further consider the first passage time using the survival probability.  The survival probability $\mathcal{S}(t)$ describing the probability that the particles do not arrive on the position $x_{ab}$ until the time $t$ reads
\begin{equation}\label{FPT1003}
\mathcal{S}(t)=\int_{-\infty}^{x_{ab}} C(x,t)dx.
\end{equation}
Let $t_f$ be the time to visit the position $x_{ab}$ for the first time. Utilizing Eq.~\eqref{FPT1003}, the PDF of $t_f$ reads
\begin{equation}\label{FPT10041212}
\varphi(t_f)=-\frac{d \mathcal{S}(t_f)}{d t_f}.
\end{equation}
From Eqs.~\ref{FPT1006} and \ref{FPT10041212}, we have
\begin{equation}\label{FPT100512ad}
\displaystyle
\begin{array}{ll}
  \varphi(t_f) &\sim -\sum_{N=1}^\infty\frac{d}{d t_f}Q_{t_f}(N)\int_{-\infty}^{x_{ab}}\chi_N^{*}(x)dx
 \\
 &=\frac{1}{2}\sum_{N=1}^\infty\frac{d}{d t_f}Q_{t_f}(N)\Big(-{\rm erfc}\left(\frac{x_{ab}-a N}{\sqrt{2} \delta \sqrt{N}}\right)+e^{\frac{2 a x_{ab}}{\delta^2}} {\rm erfc}\left(\frac{a N+x_{ab}}{\sqrt{2} \delta \sqrt{N}}\right)\Big)
\end{array}
\end{equation}
with \begin{equation*}
\frac{d}{d t_f}Q_{t_f}(N)=-\mathcal{L}_\alpha\left(\frac{N-\frac{t_f}{\langle \tau\rangle}}{(\frac{t_f}{\bar{t}})^{1/\alpha}}\right)\left(\frac{\bar{t}}{t_f}\right)^{1/\alpha}\frac{1}{\alpha t_f}
+\left(\frac{t_f}{\bar{t}}\right)^{1/\alpha}\frac{d}{d t_{f}}\mathcal{L}_\alpha\left(\frac{N-\frac{t_f}{\langle \tau\rangle}}{(\frac{t_f}{\bar{t}})^{1/\alpha}}\right).
\end{equation*}
Here ${\rm erfc}(z)$ denotes the complementary error function, i.e.,  ${\rm erfc}(z)=1-{\rm erf}(z)$ with error function ${\rm erf}(z)=\frac{2}{\sqrt{\pi}}\int_0^z\exp(-\tau^2)d\tau$. Eq. \ref{FPT100512ad} is confirmed in Fig.~\eqref{CTRWFPT} displaying a perfect match.

%
%
%
%
%
%
%

%
%
%
%

\section{Three limiting laws of the positional distribution}\label{18ctrwsect2}
Using Eq.~\eqref{aaeqfk104}, we derive three limiting laws that characterize the CTRW model under different conditions: when $S$ tends towards infinity, when $D$ approaches infinity, and when both $S$ and $D$ are  constants. More exactly, these laws include the L\'{e}vy stable distribution showing the statistics when $x-Vt\propto (St)^{1/\alpha}$, Gaussian distribution in the context of $x-Vt\propto \sqrt{Dt}$, and a general expression that encompasses all values of $D, V$ and $S$.

Taking Laplace transform on Eq.~\eqref{eq03} with respect to $t$, we have
\begin{equation}\label{18Eq101}
\mathcal{P}(k,s)=\frac{1}{s+Dk^2+ikV-S(-ik)^\alpha}.
\end{equation}
We are interested in the statistics of $x$ in the long time limit, which means that both $s$ and $k$ go  to zero in  Eq.~\eqref{18Eq101}.
Based on the above equation, the mentioned three laws will be discussed.

\subsection{L\'{e}vy stable distribution when $S\to \infty$}
\begin{figure}[htb]
  \centering
  \includegraphics[width=9cm, height=6cm]{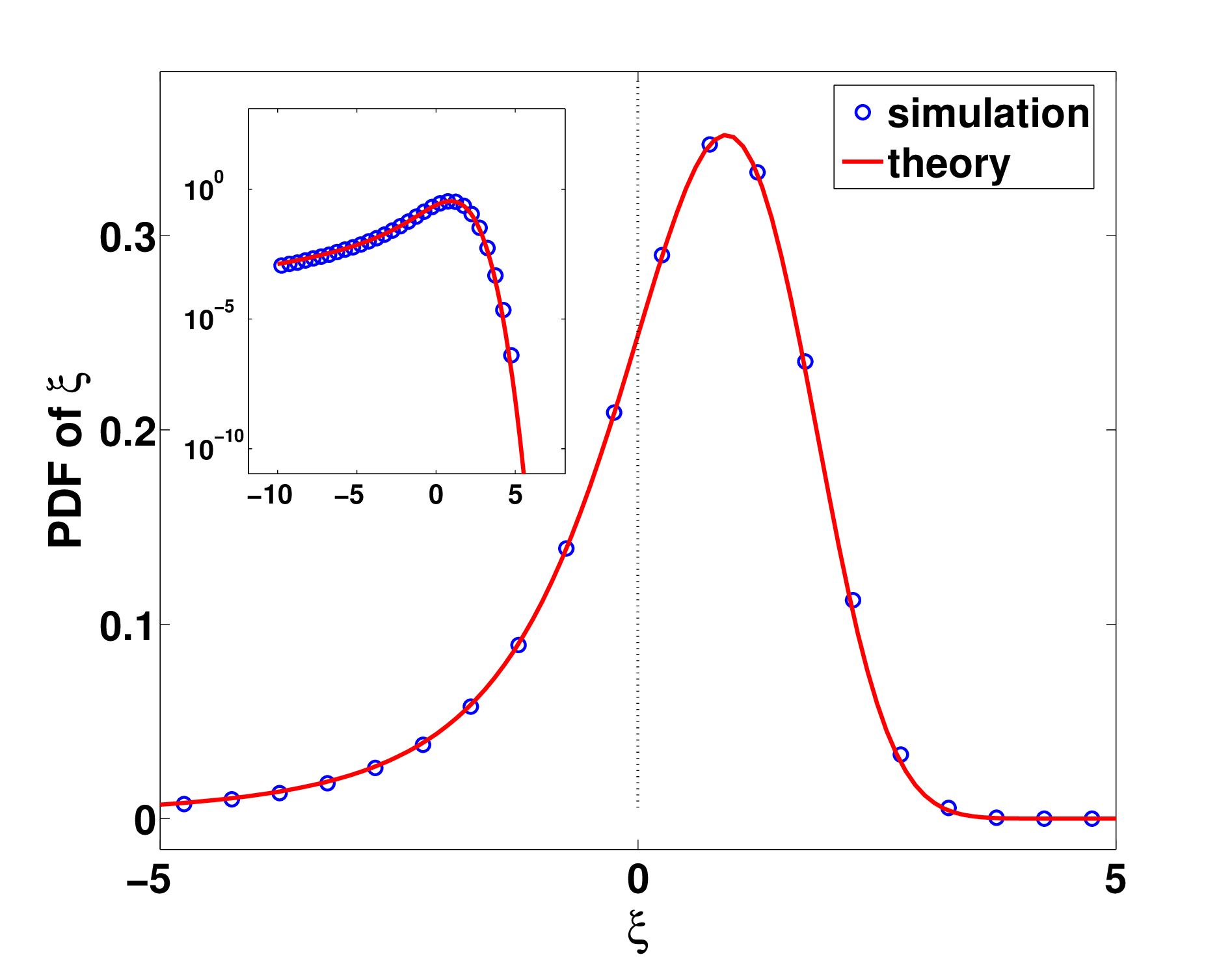}\\
  \caption{ PDF  $\mathcal{L}_{\alpha}(\xi)$ versus $\xi=(x-Vt)/(a (St)^{1/\alpha})$ for a biased CTRW model with the waiting time distribution Eq.~\eqref{ldeq3201111} and Gaussian distribution of the step length Eq.~\eqref{18eq104}. The red  line is the theoretical prediction Eq.~\eqref{18EEq102x}, showing the limit of typical fluctuations when $x-Vt$ is of the order of $t^{1/\alpha}$. The
  corresponding simulation result, represented  by  symbols  `$\circ$', is obtained by averaging $5\times10^6$ trajectories. Here the parameters are $t=1000$, $a=2$, $\sigma=0.07$, and $\tau_0=0.1$.  We will show how the limiting law Eq.~\eqref{18EEq102x} fails for a finite time $t$  or a large  $D$ and give a general law, together with a slow convergence; see Fig.~\ref{PositionEventTypical} and below discussions.
  }
\label{PositionbulkScale15}
\end{figure}

Let $S\to \infty$, Eq.~\eqref{18Eq101} yields
\begin{equation}\label{18EEq102}
\mathcal{P}(k,s)\sim \frac{1}{s+iVk-S(-ik)^\alpha},
\end{equation}
where we dropped the term $Dk^2$ due $Dk^2\ll iVk,~S(-ik)^\alpha$.
The inverse Laplace-Fourier transforms of  Eq.~\eqref{18EEq102} lead to
\begin{equation}\label{18EEq102x}
\mathcal{P}(x,t)\sim \frac{1}{(St)^{1/\alpha}}\mathcal{L}_\alpha\left(\frac{x-Vt}{(St)^{1/\alpha}}\right).
\end{equation}
Integrating Eq.~\eqref{18EEq102x} from negative infinity to positive infinity implies that $\mathcal{P}(x,t)$ is a  normalized density. See also the equivalent expression given in Eq.~\eqref{18eq109}.
This is illustrated in Fig.~\ref{PositionbulkScale15}, where the PDF of $\xi=(x-Vt)/(St)^{1/\alpha}$ with $\alpha=1.5$ is plotted. Obviously, the PDF of $\xi$ is  asymmetric with respect to $\xi$, whose two tails show two different dynamic behaviors, namely the right-hand side of the tail  tends to zero rapidly but the other one decays slowly like a power law; see the inset in Fig.~\ref{PositionbulkScale15}. More precisely, based on Eq.~\eqref{deflevystable}, we get $\mathcal{L}(\xi)\sim (-\xi)^{-1-\alpha}/\Gamma(-\alpha)$, being the same as the tail of the symmetric L{\'e}vy stable distribution, for $\xi\to -\infty$. Thus, when $q>\alpha$, the  integral $\int_{-\infty}^{\infty}|\xi|^q\mathcal{L}_{\alpha}(\xi){\rm d}\xi$ diverges. This means that Eq.~\eqref{18EEq102x} does not give any information to the MSD, which will be discussed in Appendix \ref{18ctrwSect7}.

\subsection{Gaussian distribution when $D\to \infty$}\label{finiteTimettttttt}

\begin{figure}[htb]
  \centering
  \includegraphics[width=9cm, height=6cm]{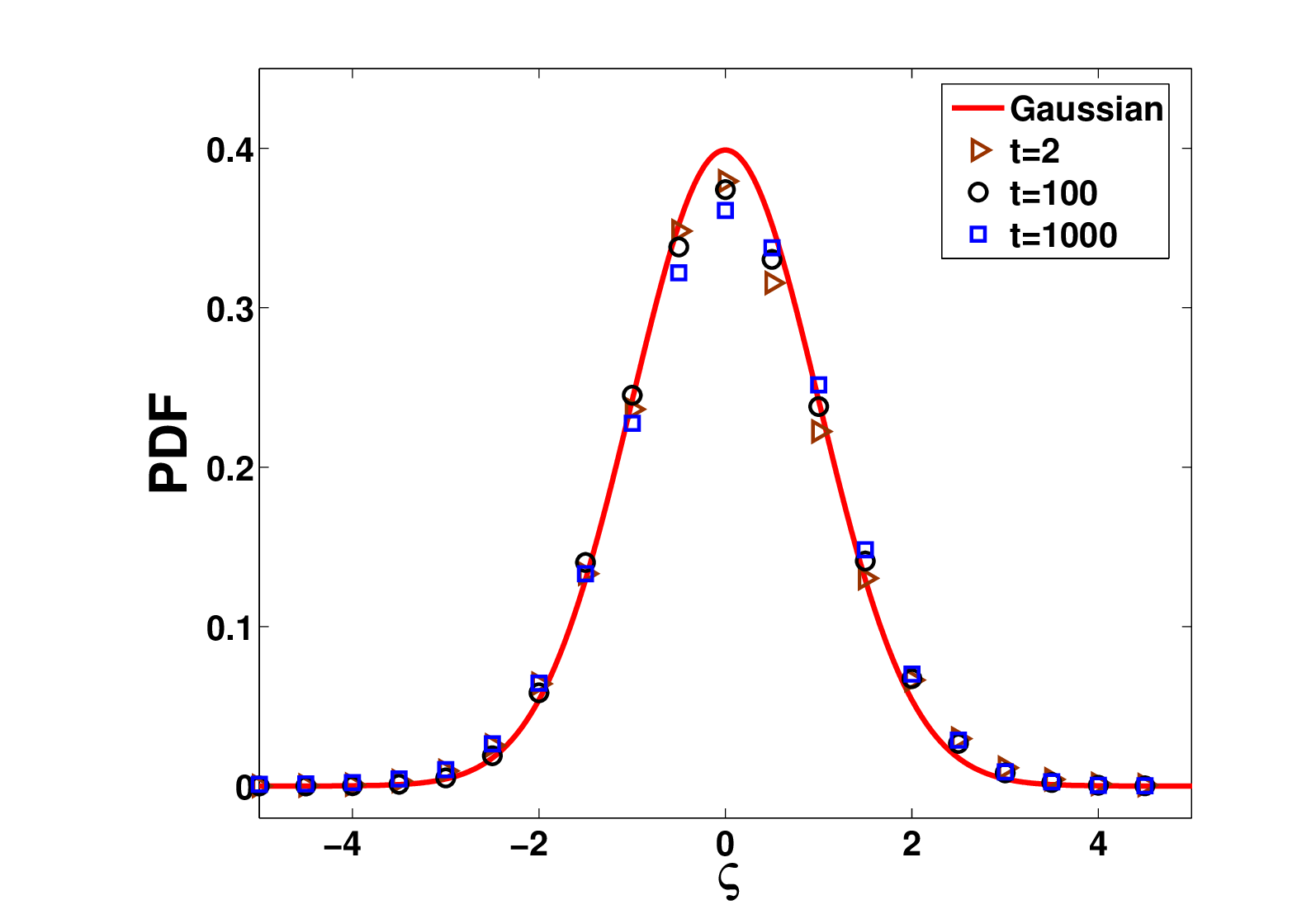}\\
  \caption{
  The dynamic of $\mathcal{P}(\varsigma)$ with scaling $\varsigma=(x-Vt)/\sqrt[]{2Dt}$ for different observation times $t$. Here we choose $\alpha=1.5$, $a=1$, $\sigma=11$, $\tau_0=0.1$, and $\langle\tau\rangle=0.3$.  These parameters suggest that the process can be approximated as nearly Gaussian, with $D$ being approximately $201.7$ and $S$ being approximately $2.3$.
  The full red line depicts the Gaussian distribution Eq.~\eqref{18eqgaussian}, which represents the behaviors when $x-Vt$ is of the order of $\sqrt{t}$. On the other hand, the symbols correspond to simulation  results obtained by averaging $10^6$ trajectories.
}
\label{Gaussian}
\end{figure}

Note that Eq.~\eqref{18EEq102x} is  independent of $D$, and when $D$ is large,  we expect this approximation to fail for a large but finite $t$. Here we focus on the case when $s+ikV\propto k^2$. As previously discussed, it is related to the linear response regime.
In this limit, according to Eq.~\eqref{18Eq101}
\begin{equation}\label{18eq110}
\mathcal{P}(k,u)\sim\frac{1}{s+Dk^2+ikV}.
\end{equation}
By inversion, using the shifting  property of the inverse Fourier transform, yields the limiting distribution of $x$.
The scaling form of $\varsigma=(x-Vt)/\sqrt[]{2Dt}$ gives the Gaussian distribution with mean zero and variance one
\begin{equation}\label{18eqgaussian}
\mathcal{P}(\varsigma)\sim \frac{1}{\sqrt{2\pi }}\exp\left(-\frac{\varsigma^2}{2}\right);
\end{equation}
see Fig.~\ref{Gaussian}.  As expected, the simulations are consistent with the theoretical result
Eq.~\eqref{18eqgaussian} for   a  large $D$. 
Recall that in Figs.~\ref{PositionbulkScale15} and  \ref{Gaussian}, we used the same observation time $t$, i.e., $t=1000$, but propagators are totally different (one is the L\'{e}vy stable law and the other is Gaussian distribution). It indicates that these laws are determined by the relationship between $D$ and $S$. The current challenge is to determine a universal form that can represent both of the mentioned scalings.


\subsection{Typical fluctuations}\label{18ctrwsect4}
\begin{figure}[htb]
  \centering
  \includegraphics[width=9cm, height=6cm]{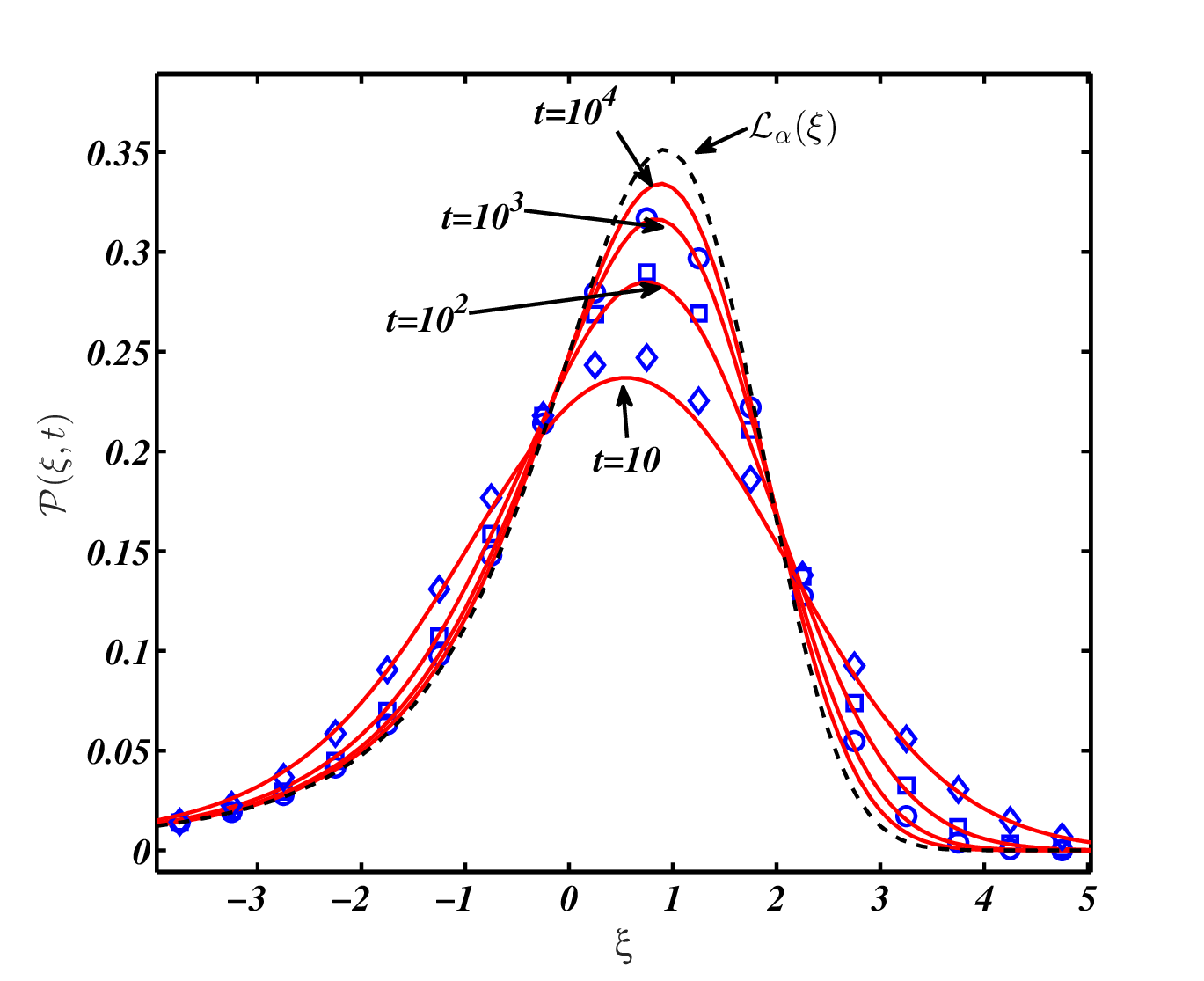}\\
  \caption{ Graph of densities $\mathcal{P}(\xi,t)$  with $\xi=(x-at/\langle\tau\rangle)/(a(t/\overline{t})^{1/\alpha})$ for different times $t$.  The  red solid lines represent the analytical scaling result obtained from Eq.~\eqref{18eqg201} and the corresponding blue symbols are simulations of $5\times10^6$ trajectories. The  dashed black line is the limiting law $\mathcal{L}_{\alpha}(\xi)$ predicted by Eq.~\eqref{18EEq102x}. Just as the figure shows, with the increase of the observation time $t$, the PDF $\mathcal{P}(\xi,t)$ tends to $\mathcal{L}_{\alpha}(\xi)$ slowly. The parameters are $\alpha=1.5$, $\tau_0=0.1$, $a=1$, $\langle\tau\rangle=0.3$ and $\sigma=1/\sqrt{2}$. See also discussions in Appendix \ref{xxxxxB}.
}
\label{PxtWithTime}
\end{figure}

\begin{figure}[htb]
  \centering
  \includegraphics[width=9cm, height=6cm]{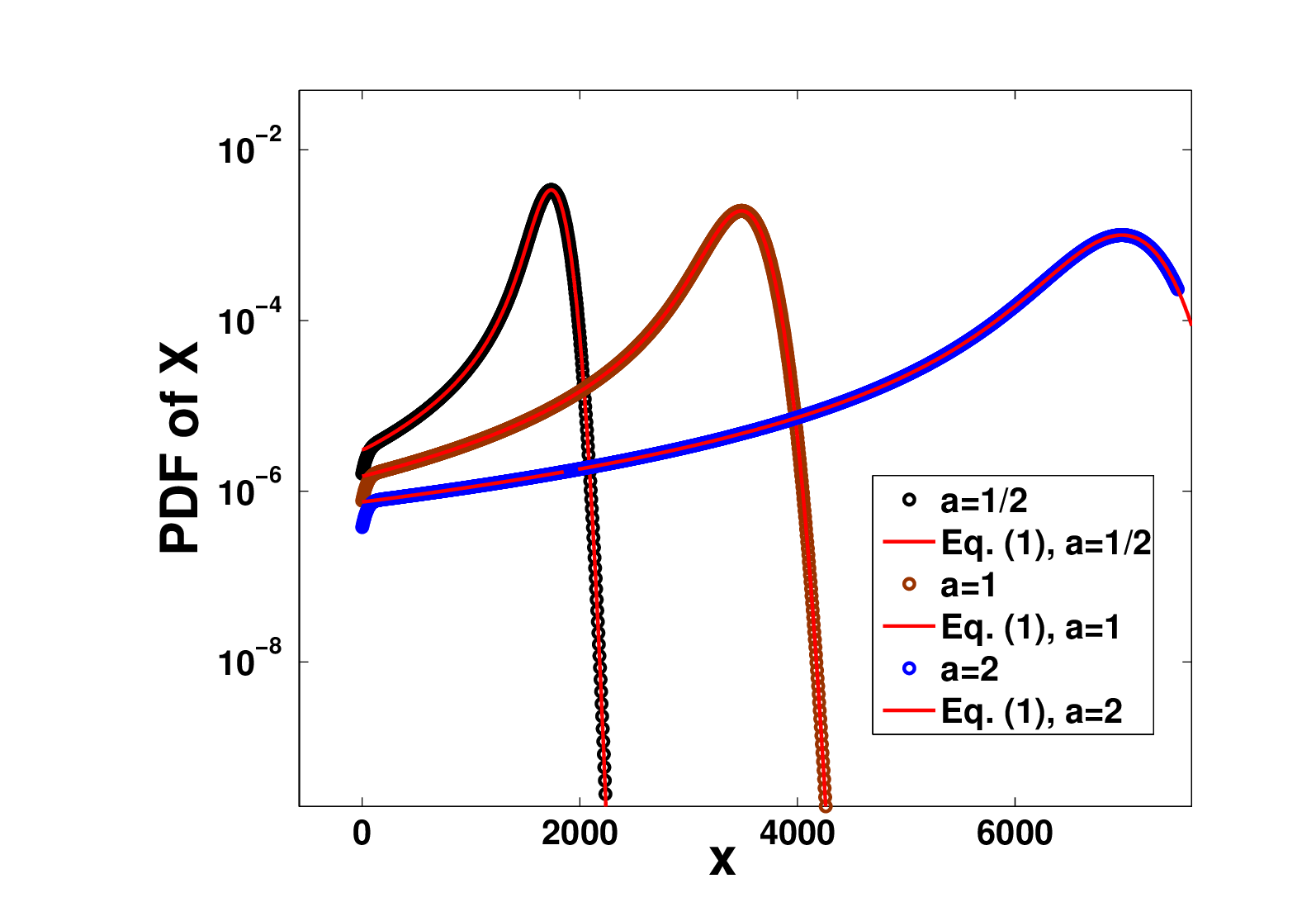}\\
  \caption{Comparison between Eq.~\eqref{18eq504} and Eq.~\eqref{18eqg201} for different forces. The symbols are the plot of Eq.~\eqref{18eq504} and the solid lines  describe the solution of Eq.~\eqref{aaeqfk104} or  Eq.~\eqref{18eqg201}. Here we choose $t=1000$, $t_0=0.1$, $\alpha=1.5$, and $\delta=1$.
}
\label{DifferenceOfTypical}
\end{figure}

We have discussed that under certain conditions  the positional distribution follows  L\'{e}vy stable distribution Eq.~\eqref{18EEq102x} or Gaussian distribution Eq.~\eqref{18eqgaussian}, determined by the relation between $D$ and $S$. Therefore, it would be beneficial to acquire a universal law that applies to the aforementioned scenarios. Based on  Eq.~\eqref{18Eq101}, we have
\begin{equation}\label{18eqg201}
 \mathcal{P}(x,t) = \frac{1}{\sqrt{4\pi D }S^{1/\alpha}t^{1/\alpha+1/2}} \int_{-\infty}^{\infty}\mathcal{L}_\alpha\left(\frac{y}{(St)^{1/\alpha}}\right)\exp\left(-\frac{(x-Vt-y)^2}{4Dt}\right)dy.
\end{equation}
We denote Eq.~\eqref{18eqg201}, i.e., the solution of Eq.~\eqref{aaeqfk104}, as typical fluctuations describing the central of the positional distribution. As mentioned before, in the long time limit, Eq.~\eqref{18eqg201} leads to the L{\'e}vy stable distribution Eq.~\eqref{18EEq102x}. Figure  \ref{PxtWithTime} demonstrates the validity of Eq. \eqref{18eqg201} for various values of $t$.
For comparison, we also plot the solution Eq.~\eqref{18eq504} in Fig.~\ref{DifferenceOfTypical}. Eq.~\eqref{18eq504} agrees with Eq.~\eqref{18eqg201} for large $x$, it rapidly approaches zero as $x$ approaches zero, whereas Eq.~\eqref{18eqg201} exhibits a cutoff at the tail. Therefore, the primary difference between the two equations lies in small values of $x$.

To  summarize, typical fluctuations Eq.~\eqref{18eqg201},  giving the information when $x-Vt\propto t^{1/\alpha}$, are valid for numerous  $D$, $V$, $S$ and $t$. One may wonder ``why Eq.~\eqref{18eqg201} is more useful than Eq.~\eqref{18eq109}?''. The reason is as follows:  Mathematically,  Eq.~\eqref{18eq109} is valid under the condition that   $gg(y,t,\sigma,a,\alpha)\to 0$. This  implies that the observation time $t$ should be  especially long, for example  when $a$ is small and $\sigma\to \infty$. While, Eq.~\eqref{18eqg201} avoids this, this means that the condition reduces to $\langle N\rangle\propto t/\langle \tau\rangle$.

\section{Conclusion and discussion}\label{18ctrwSect8}

In the context of the CTRW model, if the displacement follows a narrow distribution with a non-zero mean and the waiting time has an infinite mean, the fractional operator of the diffusion equation is linked to the Riemann-Liouville time derivative, as discussed in Refs. \cite{Metzler2000random,Barkai2001Fractional}. However, nearly two decades later, the fractional-space advection-diffusion equation was developed for situations in which waiting times possess a finite mean and an infinite variance, together with the mentioned displacements \cite{Wang2020Fractional}.
As an extension of work \cite{Wang2020Fractional}, here we demonstrated further evidence of Eq.~\eqref{aaeqfk104}, gave more applications, and showed limiting laws of the kinetic equation, through our in-depth explanations. At the same time, theoretical predictions are checked by simulations of the CTRW model.

An issue here is that for L{\'e}vy flights, with a fat-tailed distribution of jump lengths, the method of images was shown to lead to wrong results, due to the non-local behavior \cite{Chechkin2003First,Sokolov2004Non,Palyulin2019First}. One may wonder, whether or not the method of images will work for our process which does not include any fat-tailed jump length distribution. We show that generally, it does not work, unless for the case when the bias is small, then the image method is a valid approximation. For that, a subordination method was used. This method holds in general beyond the first passage problem, and it is also very different if compared to the subordination method used for $0<\alpha<1$. In practice we use the image method on the conditional probability $\chi_{N}(x)$, instead of the positional distribution $\mathcal{P}(x,t)$. In particular, for a free particle, we assume Eq.~\eqref{18eq502add1} has a free boundary condition. In that sense, $\chi^{*}_N(x)$ Eq.~\eqref{FPT1005} reduces to $\chi_N(x)$ given in Eq.~\eqref{GaussianN}.
The mentioned strategy can also be extended to a much more general case, for example, the diffusion of particles on a finite domain. For applications, we also discussed the two-dimensional diffusion equation and breakthrough through curves with time-dependent bias and variance.

%
%
%

We demonstrated that the solution of Eq.~\eqref{aaeqfk104} is obtained by convolving the Gaussian distribution with the L{\'e}vy stable distribution, see Eq.~\eqref{18eqg201}. Here we denote the solution of Eq.~\eqref{aaeqfk104} as typical fluctuations of CTRW describing a wide range of $D$, $V$, and $S$.
In the language of the biased CTRW, the L{\'e}vy stable distribution stems from the PDF of the number of renewals, and the Gaussian distribution comes from the conditional distribution $\chi_N(x)$. When the bias is strong, i.e., $S\to\infty$, the diffusion term in Eq.~\eqref{aaeqfk104} can be ignored and the solution is an asymmetric L{\'e}vy stable distribution Eq.~\eqref{18EEq102x}. For the CTRW model, this limit is achieved when the ratio $\sigma^2/a^2$ is finite and $t\to \infty$ or $\sigma^2/a^2\to 0$ with a finite $t$. On the contrary, when $S\to 0$, the spreading packet follows Gaussian distribution. It indicates that when $\sigma^2/a^2 \gg 1$, the well-known  Eq.~\eqref{18EEq102x} is not useful unless $t$ is extremely long. Furthermore, we provide a parameter, denoted as $\overline{\sigma}^2$, in Eq.~\eqref{sigmaneq} to characterize the convergence.

There are still unanswered questions that require addressing.
For example, in the simulation of particle trajectories, generally, we first generate waiting times and displacements. Then, upon the completion of the total observation time, denoted as $t$, we obtain positional statistics.
However, when the observation time is long and the number of realizations is large, such as $10^7$, running codes may require several days to complete. The problem now is whether we can generate particle statistics based on Eq.~\eqref{aaeqfk101}. More precisely, we first generate a variable drawn from the L{\'e}vy stable distribution $\mathcal{L}(y)$, and then use it to generate the position at time $t$ according to a Gaussian distribution. This method seems fine if we are only interested in the positional distribution, but it fails as expected when we focus on the MSD. Solving this problem, i.e. formulating Langevin paths,  is a matter that will be considered in the future.

\section*{Acknowledgments}
W.W. is supported by the National Natural Science Foundation of China under Grant No.
12105243 and the Zhejiang Province Natural Science Foundation LQ$22$A$050002$.   E.B. acknowledges the Israel Science Foundations grant $1614/21$.

\appendix
\begin{appendices}

\section{Additional discussions on Fig.~\ref{PositionEventTypical}}\label{xxxxxB}
From simulations of typical fluctuations in Fig. \ref{PositionEventTypical}, we can see that
we have $\overline{\sigma}^2\simeq 0.1115\sigma^2/a^2$ defined by Eq.~\eqref{sigmaneq} with $t=1000$, $\alpha=1.5$, and $\tau_0=0.1$.  In the particular case $\sigma=0.1/\sqrt{2}$, from Eq.~\eqref{sigmaneq} we get a small $\overline{\sigma}^2\simeq 5.57\times 10^{-4}$, this is the reason why Eq.~\eqref{18eq504} tends to the limit theorem Eq.~\eqref{18eq109} or \eqref{18EEq102x} as shown by the black dashed line in Fig.~\ref{PositionEventTypical}.
However, when $\sigma =16/\sqrt{2}$, $\sigma^2/a^2 =128$  and then $\overline{\sigma}^2=14.27$, which is certainly not a small number if compared with $5.57\times 10^{-4}$. Thus even though the average of renewals $\langle N(t) \rangle\sim t/\langle\tau\rangle=1000/0.3\approx 3333$, we cannot say that
the  limit of $t\to \infty$ is reached, and indeed,  just as the bottom line in Fig. \ref{PositionEventTypical} shows, we see for this case nearly Gaussian behavior. For asymmetry  breaking properties, one choice is to consider $\overline{\sigma}^2$. This means that when $\overline{\sigma}^2\to 0$, the corresponding density is asymmetric. While, for a large $\overline{\sigma}^2$, the opposite situation emerges.\\

\section{Fractional moments}\label{18ctrwSect7}

\begin{figure}[htb]
  \centering
  \includegraphics[width=9cm, height=6cm]{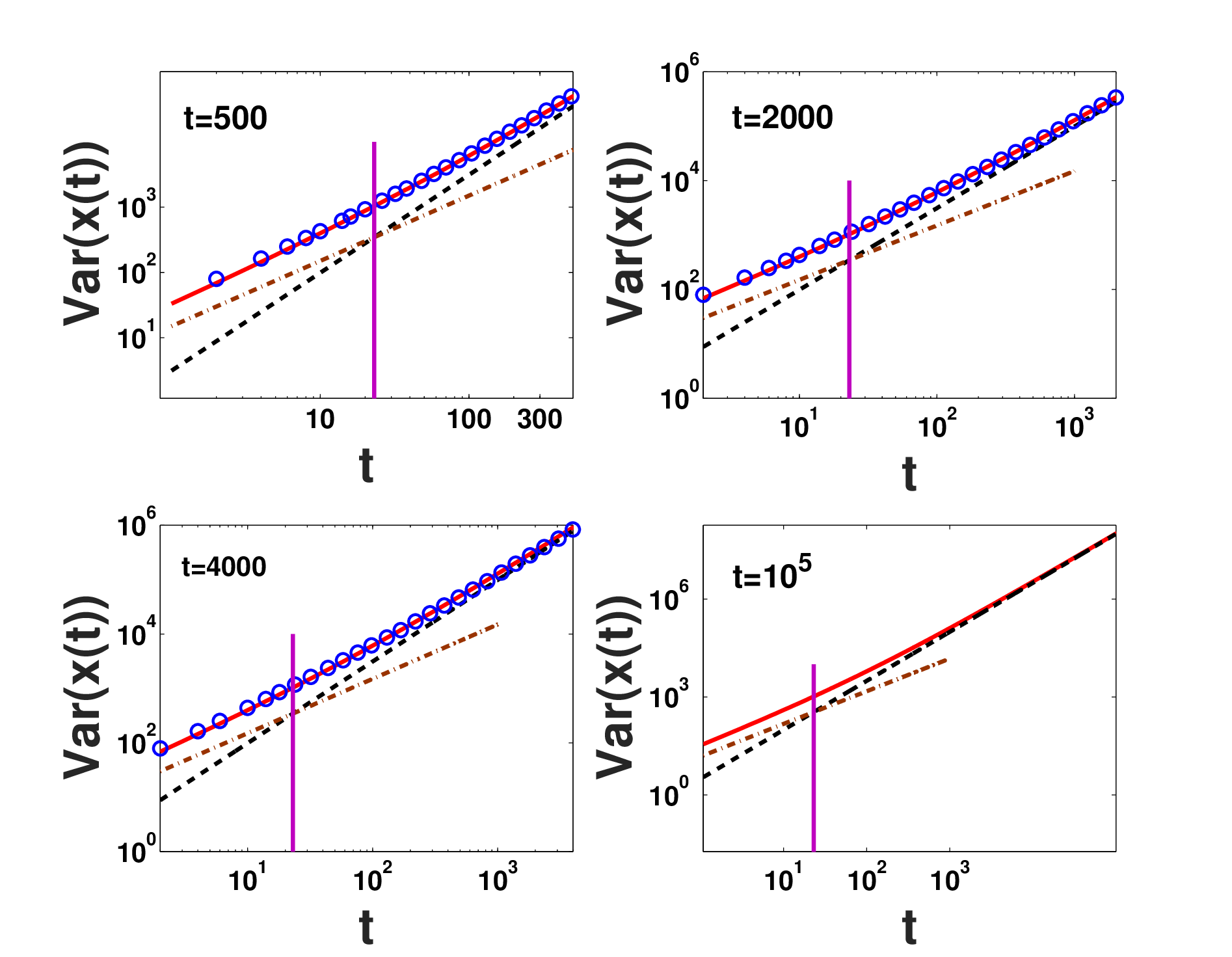}\\
  \caption{ MSD for  observation time $t=500$ [left top panel], $t=2000$ [right top panel], $t=4000$ [right bottom panel], $t=10^5$ [left bottom panel], where the parameters are $\alpha=1.5$, $\sigma=3/\sqrt{2}$, $a=1$, and $\tau_0=0.1$. The solid lines are the theoretical result  Eq.~\eqref{18msd101} and the dashed curves are the dominating term of Eq.~\eqref{18msd101}, i.e., $(2a^2\tau_0^\alpha)/((2-\alpha)(3-\alpha)\langle\tau\rangle^3)t^{3-\alpha}$. The symbols are simulations obtained by averaging $10^4$ realizations.}
\label{moments2force}
\end{figure}

As mentioned before, Eq.~\eqref{aaeqfk104} fails to estimate MSD of the CTRW model. Recently, we give a way to compute the MSD using infinite densities \cite{Wang2019Transport}. Here we show that infinite density frameworks can also used to calculate the fractional moments in an asymptotic sense.
For the sake of completeness, first let us concentrate on the variance of  the walk's displacement, called MSD, which is a measure of the deviation of the position of the particle with respect to the position over time. Suppose that $x=0$ is the initial position of particles and $N$ is the number of renewals between $0$
and $t$. Thus, the first and the second order moment of the position are given by
\begin{equation}
\langle x\rangle=\left\langle\sum_{j=1}^{N}\chi_i\right\rangle
\end{equation}
and
\begin{equation}
\langle x^2\rangle=\left\langle \left(\sum_{j=1}^{N} \chi_i\right)^2\right\rangle,
\end{equation}
respectively. Here $\chi_i$ denotes the displacement of the particle in the $i$-th jump and $ \langle \cdot \rangle $ describes the average over $\chi_1, \chi_2, \dots, \chi_N$. Therefore, the variance is
\begin{equation}\label{18ctrwfm101}
V\!ar(x)=\langle x^2\rangle-\langle x\rangle^2
=\langle N\rangle (\langle\triangle x^2\rangle-\langle\triangle x\rangle^2)+(\langle N^2\rangle-\langle N\rangle^2)\langle\triangle x\rangle^2,
\end{equation}
where we assumed that $\chi_i$ are IID random variables, and $\langle\triangle x\rangle$ and
$\langle\triangle x^2\rangle$ correspond to the first and the second  order moments of $\chi_i$, respectively.  Let us take the step length distribution as Eq.~\eqref{18eq104},  thus we have $\langle\triangle x\rangle=a$ and $\langle\triangle x^2\rangle=a^2+\sigma^2$. To obtain $V\!ar(x)$, we need to calculate the first and the second moments of $N$. Using the previous results given in Ref.~\cite{Godreche2001Statistics}, we have
\begin{equation}\label{gg18msd101}
\langle \widehat{N}(s)\rangle=\frac{\widehat{\phi}(s)}{s(1-\widehat{\phi}(s))}
\end{equation}
and
\begin{equation}\label{gg18msd102}
\langle \widehat{N}^2(s)\rangle=\frac{\widehat{\phi}(s)(1+\widehat{\phi}(s))}{s(1-\widehat{\phi}(s))^2}.
\end{equation}
Utilizing Eqs.~\eqref{gg18msd101} and \eqref{gg18msd102}, Eq.~\eqref{18ctrwfm101} gives
\begin{equation}\label{18msd101}
 V\!ar(x)\sim \frac{2a^2\tau_0^\alpha}{(2-\alpha)(3-\alpha)\langle\tau\rangle^3}t^{3-\alpha}+\sigma^2\frac{t}{\langle\tau\rangle},
\end{equation}
which is plotted in  Fig.~\ref{moments2force}. The term $t^{3-\alpha}$ becomes the leading term of $V\!ar(x)$ as we increase of the observation time $t$; See also discussions in Ref.~\cite{Wang2019Transport}. When the observation time is not very large, the linear term $\sigma^2t/\langle\tau\rangle $ wins since the spreading packet nearly follows Gaussian distribution; see the dash-dotted line in Fig.~\ref{moments2force}.
For a fixed observation time $t$,
we can find an interesting phenomenon depicting the competition  between anomalous diffusion and normal diffusion, which is determined by $a$, $\sigma$, and $t$. According to Eq.~\eqref{18msd101}, the transition point is
\begin{equation}
t^*\sim \left((2-\alpha)(3-\alpha)\frac{\sigma^2\langle\tau\rangle^2}{2a^2\tau_0^\alpha}\right)^{\frac{1}{2-\alpha}};
\end{equation}
see the magenta vertical lines in Fig.~\ref{moments2force}. When $|a|\to 0$, $t^*$ goes to infinity. It indicates that when the bias is weak, the process needs a long observation time $t$ to exhibit anomalous behaviors. On the other hand, when $a$ approaches infinity, the value of $t^*$ tends to zero, causing the transition from normal diffusion to anomalous diffusion rapidly.

When $q=4$, the corresponding moment is
\begin{equation}\label{18msd102}
\langle (x-\langle x\rangle)^4\rangle=\langle x^4\rangle+
6\langle x\rangle^2\langle x^2\rangle-4\langle x\rangle\langle x^3\rangle-3\langle x\rangle^4
\end{equation}
Using Eq.~\eqref{18eq500}, the integer order moments are easy to get. Then substituting them into Eq.~\eqref{18msd102} gives
\begin{equation}\label{18msd103}
\langle (x-\langle x\rangle)^4\rangle\sim a^4\langle (N-\langle N\rangle)^4\rangle+6 a^2\sigma^2(\langle N^3\rangle+\langle N\rangle^3-2\langle N\rangle\langle N^2\rangle)
\end{equation}
Now  the calculation  of  $\langle  (x-\langle x\rangle)^4\rangle$ is transformed into the moments of the number of renewals. After some simple arithmetics, we have

\begin{equation}\label{18msd103a}
\displaystyle{
          \begin{array}{ll}
            \langle (x-\langle x\rangle)^4\rangle &\sim\frac{4a^4b_\alpha}{|\Gamma(1-\alpha)|\langle \tau\rangle^5(5-\alpha)(4-\alpha)}t^{5-\alpha}
+\frac{6a^2\sigma^2b_\alpha}{\langle \tau\rangle^4}\Big(\frac{18}{\Gamma(5-\alpha)}-\frac{8}{\Gamma(4-\alpha)}+\frac{1}{\Gamma(3-\alpha)}\Big)t^{4-\alpha}
\\
&~~~+ \frac{12a^4b_{\alpha}}{\langle \tau\rangle^4}\Big(\frac{16}{\Gamma(5-\alpha)}-\frac{4}{\Gamma(4-\alpha)}+\frac{1}{\Gamma(3-\alpha)}\Big)t^{4-\alpha}.
          \end{array}}
\end{equation}

%
%
%

Our next step is to explore $q$ order absolute  moments of $\epsilon= x-at/\langle \tau\rangle$, defined by
\begin{equation}
\langle|\epsilon|^q\rangle=\int_{-\infty}^{\infty}|\epsilon|^q P(\epsilon,t){\rm d}\epsilon
\end{equation}
with $q>0$.
Let us start from the calculation of
low order moments, i.e., $q<\alpha$. The low order moments can be obtained by the limit of  typical fluctuations Eq.~\eqref{18eq109}, which reads
\begin{equation}\label{18msd104}
\langle|\epsilon|^q\rangle\sim a^q(t/\overline{t})^{q/\alpha}\int_{-\infty}^{\infty}y^q\mathcal{L}_{\alpha}(y){\rm d}y.
\end{equation}
When $q<\alpha$, the integral $\int_{-\infty}^{\infty}y^q\mathcal{L}_{\alpha}(y){\rm d}y$ is a finite number which can be  evaluated  by the asymptotic behaviors of L{\'e}vy distribution $\mathcal{L}_{\alpha}(y)$. We can check that for $0<q<\alpha$, $\langle|\epsilon|^q\rangle\sim t^{q/\alpha}$ is always sublinear in $t$.
Here we would like to consider two special cases: for $q\to 0$,
the zeroth moment of the variable $\epsilon$  is, evidently, one; the other case is the Gaussian limit,  the  linear time dependence of the MSD is recovered, namely $\lim_{\alpha,q \to 2}\langle|\epsilon|^q\rangle\sim t$.
For $q>\alpha$, Eq.~\eqref{18msd104} diverges. It indicates that the normalized density Eq.~\eqref{18eq109} can not give a valid prediction for high order moments; see Eq.~\eqref{18ctrwfm101}.

Now we consider high-order moments. As mentioned before, for this case, typical fluctuations does not work.
In \cite{Wang2019Transport}, we demonstrate the existence of an additional limiting law when the second time scale is taken into account. This law, denoted as Eq.~\eqref{rareeq204}, characterizes the scaling behavior when $x-at/\langle\tau\rangle$ is of the order of  $t/\langle\tau\rangle$. Namely,
the density of $\eta=(x-at/\langle \tau\rangle)/(at/\langle\tau\rangle)$ is
\begin{equation}\label{rareeq204}
P(\eta,t)\sim \frac{b_\alpha t^{1-\alpha}}{\Gamma(-\alpha)\langle\tau\rangle}(-\eta)^{-\alpha-1}\Big(1-\frac{1-\alpha}{\alpha}\eta\Big)
\end{equation}
with  $-1<\eta<0$. See numerical simulations and discussions in Ref. \cite{Wang2019Transport}.

As mentioned,
this situation for $q=2$ is especially important in physical and biological  experiments \cite{Barkai2012Strange,Metzler2000random}. After rescaling Eq.~\eqref{rareeq204} by $\epsilon= x-at/\langle \tau\rangle$, it can be seen that $|\epsilon|^q$ is integrable with respect to this non-normalized state. So this can be used to get high order moments, i.e.,
\begin{equation}\label{18msd105}
\langle|\epsilon|^q\rangle\sim \frac{a^qq \tau_0^\alpha}{\langle\tau\rangle^{q+1}(q-\alpha)(q-\alpha+1)}t^{q+1-\alpha}.
\end{equation}
This demonstrates that, in the long time regime, the asymptotic distribution of $\epsilon$ is broad with a slowly decaying tail; see Eq.~\eqref{rareeq204}. Note that when $t$ is not  sufficient large, the correction term is necessary to be added since the packet of particles is Gaussian; see left top panel in Fig.~\ref{moments2force}.

To summarize, the limiting law  Eq.~\eqref{18eq109} does describe the low order  moments. The rare fluctuations, predicted by Eq.~\eqref{rareeq204}, give the information to the events when $\epsilon$ is of the order of $t$. In the long time limit, the results of $q$ order moments can be generalized to
\begin{equation}
         \langle|\epsilon|\rangle\sim t^{\varrho(q)},
\end{equation}
where  $\varrho(q)=q/\alpha$ if $q<\alpha$, and $\varrho(q)=q+1-\alpha$ if $q>\alpha$.

\section{Plot of concentration with a weak bias}\label{CTRWAPPENc}
\begin{figure}[h]
  \centering
  \includegraphics[width=9cm, height=6cm]{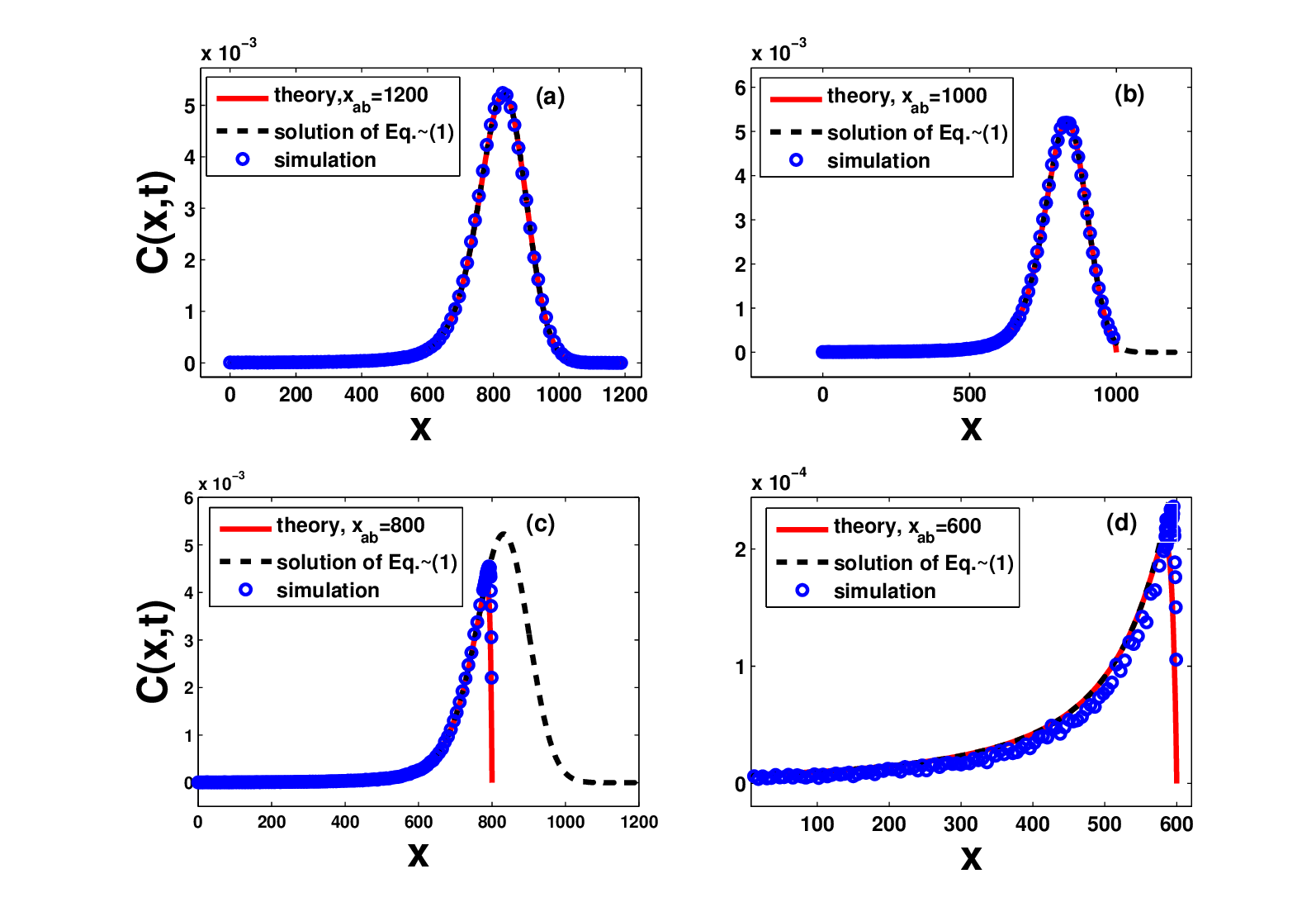}\\
  \caption{Plot of concentration Eq.~\eqref{FPT1004} for different absorbing boundary conditions, i.e., $x_{ab}=1200, 1000, 800, 600$ for (a), (b), (c), (d) respectively. The parameters for CTRW simulations are $10^7$ realizations, $t=800$, $\alpha=1.5$, $\sigma=1$, $a=0.3$, and $\tau_0=0.1$. Red solid lines demonstrate the theoretical predication  Eq.~\eqref{FPT1004} with $\mathcal{P}(x,t)$ being the solution of Eq.~\eqref{aaeqfk104} and dashes lines are the solution of Eq.~\eqref{aaeqfk104}, i.e., without absorbing conditions,  for comparison.}
\label{ObPXT}
\end{figure}

We plot Eq.~\eqref{FPT1004} obtained from the image method for a weak bias; see Fig.~\ref{ObPXT}. For this case, the asymmetric term $S$ loses its role and the process is nearly normal. As mentioned in the main text, if the bias is strong, Eq.~\eqref{FPT1004} breaks down.

\end{appendices}
\bibliography{wenxian}

\end{document}